\documentstyle[draft,epsf]{mn}

\def\gsim{\mathrel{\raise0.35ex\hbox{$\scriptstyle >$}\kern-0.6em 
\lower0.40ex\hbox{{$\scriptstyle \sim$}}}}
\def\lsim{\mathrel{\raise0.35ex\hbox{$\scriptstyle <$}\kern-0.6em 
\lower0.40ex\hbox{{$\scriptstyle \sim$}}}}
\def\hi{{\rm H\,{\sc i} }}
\def\hii{{\rm H\,{\sc ii} }}
\def\tv96{{\rm TVPHW}}

\title{The stellar populations of spiral galaxies}

\author[Bell \& de Jong]
{
Eric F. Bell$^1$ \& Roelof S. de Jong$^{1,2}$\thanks{Hubble Fellow} \\
$^1$ Department of Physics, University of Durham, 
South Road, Durham DH1 3LE, UK\\
$^2$ Steward Observatory, University of Arizona, 
949 N. Cherry Ave., Tucson, Arizona, 85719, USA
}

\begin{document}
\date{\fbox{{\sc Submitted to MNRAS}: \today}}

\label{firstpage}

\maketitle

\begin{abstract}
We have used a large sample of low-inclination spiral galaxies
with radially-resolved optical and near-infrared photometry to 
investigate trends in star formation history with radius as a
function of galaxy structural parameters.  
A maximum likelihood method was used to match all the available 
photometry of our sample to the colours predicted
by stellar population synthesis models.
The use of simplistic star formation histories, 
uncertainties in the stellar population models and 
regarding the importance of dust
all compromise the absolute ages and metallicities
derived in this work, however our conclusions are robust
in a relative sense.  We find that most spiral
galaxies have stellar population gradients, in the sense
that their inner regions are older and more metal rich than 
their outer regions.  Our main conclusion is that 
the surface density of a galaxy drives its star formation history, 
perhaps through a local density dependence in the star formation law.
The mass of a galaxy is a less important parameter; the age of a galaxy
is relatively unaffected by its mass, however the metallicity 
of galaxies depends on both surface density and mass. 
This suggests that galaxy mass-dependent feedback 
is an important process in the 
chemical evolution of galaxies.  In addition,  
there is significant cosmic scatter suggesting
that mass and density may not be the 
only parameters affecting the star formation history of a galaxy.
\end{abstract}


\section{Introduction}

It has long been known that there are systematic trends in the
star formation histories (SFHs) of spiral galaxies;
however, it has been notoriously difficult
to quantify these trends and directly relate them to 
observable physical parameters.
This difficulty stems from the age-metallicity degeneracy:
the spectra of composite stellar populations are 
virtually identical if the percentage change in age or metallicity (Z)
follows $\Delta{\rm age}/\Delta{\rm Z} \sim 3/2$ \cite{w94}.
This age-metallicity degeneracy is split only for certain
special combinations of spectral line indices or for 
limited combinations of broad band colours 
(Worthey 1994; de Jong 1996c; hereafter dJ{\sc iv}).
In this paper, we use a combination of optical and near-infrared (near-IR)
broad band colours to partially break the age-metallicity degeneracy and
relate these changes in SFH with a galaxy's observable physical parameters.

The use of broad band colours is now a well established
technique for probing the SFH of galaxy populations.
In elliptical and S0 galaxies there is a strong
relationship between galaxy colour and absolute magnitude 
\cite{sandage1978,larson1980,ble92,ale}: this is the so-called
colour-magnitude relation (CMR).  This correlation, which is
also present at high redshift in both the cluster and the field
\cite{ellis1997,stanford1997,kodama1998b}, 
seems to be driven by a metallicity-mass
relationship in a predominantly old stellar population 
\cite{ble92,kodama1998a,bower1998}.
The same relationship can be explored for spiral galaxies:  Peletier
and de Grijs \shortcite{peletier1998} determine a dust-free
CMR for edge-on spirals.  They find a tight
CMR, with a steeper slope than for elliptical and S0 galaxies.
Using stellar population synthesis models, they conclude that 
this is most naturally interpreted as indicating 
trends in both age and metallicity with magnitude, with 
faint spiral galaxies having both a younger age and
a lower metallicity than brighter spirals. 

In the same vein, radially-resolved colour information can be used to
investigate radial trends in SFH.  In dJ{\sc iv}, radial colour
gradients in spiral galaxies were found to be common and were found to
be consistent predominantly with the effects of a stellar age gradient. 
This conclusion is supported by Abraham et al.\
\shortcite{abraham1998}, who find that dust and metallicity gradients
are ruled out as major contributors to the radial colour trends observed in
high redshift spiral galaxies.  Also in dJ{\sc iv},
late-type galaxies were found to be on average younger and
more metal poor than early-type spirals.  As late-type spirals
are typically fainter than early-type galaxies
(in terms of total luminosity and surface 
brightness; de Jong 1996b), these
results are qualitatively consistent with those of Peletier \& de Grijs
\shortcite{peletier1998}.  
However, due to the
lack of suitable stellar population synthesis models at that time, the
trends in global age and metallicity in the sample were difficult to
meaningfully quantify and investigate. 

In this paper, we
extend the analysis presented in dJ{\sc iv} by using the new stellar
population synthesis models of Bruzual \& Charlot (in preparation) 
and Kodama \& Arimoto (1997), and by augmenting
the sample with low-inclination galaxies from the Ursa Major cluster
from Tully et al.\ (1996; \tv96 hereafter)
and low surface brightness galaxies from the sample of 
Bell et al.\ (1999; hereafter Paper {\sc i}).  
A maximum likelihood method is used to match all the available 
photometry of the sample galaxies to the colours predicted by these
stellar population synthesis models, allowing investigation of 
trends in galaxy age and metallicity as a function of
local and global observables.

The plan of the paper is as follows.  In section \ref{sec:data}, we
review the sample data and the stellar population synthesis and dust 
models used in this analysis.  In section \ref{sec:max}, we present the
maximum-likelihood fitting method and review the construction of the
error estimates.  In section \ref{sec:res}, we present the results and
investigate correlations between our derived ages and 
metallicities and the galaxy
observables.  In section \ref{sec:disc} we discuss 
some of the more important correlations and the limitations of
the method, compare with literature data, and discuss 
plausible physical mechanisms for generating some of the correlations 
in section \ref{sec:res}.
Finally, we review our main conclusions in section \ref{sec:conc}.
Note that we homogenise the distances
of Paper {\sc i}, dJ{\sc iv} and \tv96
to a value of $H_0 = 65$ kms$^{-1}$. 

\section{The data, stellar population and dust models} \label{sec:data}

\subsection{The data}

In order to investigate SFH trends in spiral galaxies, it is 
important to explore as wide a range of galaxy types, magnitudes, 
sizes and surface brightnesses as possible.  Furthermore, high-quality
surface photometry in a number of optical and at least one near-IR
passband is required.  Accordingly, we include the samples of 
de Jong \& van der Kruit (1994; dJ{\sc i} hereafter), \tv96 and Paper
{\sc i} in this study to cover a wide range of galaxy luminosities, 
surface brightnesses and sizes.
The details of the observations and data reduction can be found in the
above references;  below, we briefly 
outline the sample properties and analysis techniques.

The sample of undisturbed spiral galaxies described in dJ{\sc i} was
selected from the UGC \cite{ugc} to have red diameters
of at least 2 arcmin and axial ratios greater than 0.625.  The sample
presented in Paper {\sc i} was taken from 
a number of sources
\cite{deblok1995,deblok1996,oneil1997a,oneil1997b,spray95,esolv,dji}
to have low estimated blue central surface brightnesses 
$\mu_{B,0} \ga $ 22.5 
mag$\,$arcsec$^{-2}$ and diameters at the 25 B mag$\,$arcsec$^{-2}$
isophote larger than 16 arcsec. 
\tv96 selected their galaxies from the Ursa Major Cluster 
(with a velocity dispersion of only 148 km\,s$^{-1}$, it is the least
massive and most spiral-rich nearby cluster):
the sample is complete down to at least a $B$ band magnitude
of $\sim$ 14.5 mag, although their sample includes many galaxies fainter
than that cutoff.
We take galaxies from the three samples
with photometric observations 
(generously defined, with maximum allowed calibration errors of 0.15)
in at least two optical and one near-IR passband, axial ratios greater
than 0.4, and accurate colours (with errors in a single passband due to 
sky subtraction less than 0.3 mag)
available out to at least 1.5 $K$-band disc scale lengths.
For the Ursa Major sample, we applied an additional selection criterion
that the $K'$ band disc scale length must exceed 5 arcsec, to allow
the construction of reasonably robust central colours (to avoid 
the worst of the effects of mismatches in seeing between the different
passbands).
This selection leaves a sample of 64 galaxies from dJ{\sc i}
(omitting UGC 334, as we use the superior data from 
Paper {\sc i} for this galaxy: note that we show the
two sets of colours for UGC 334 connected with 
a dotted line in Figs.\,\ref{fig:colcolmag}--\ref{fig:colcolscal}
to allow comparison), 
23 galaxies from Paper {\sc i}, and 34 galaxies from 
\tv96 (omitting NGC 3718 because of its highly peculiar morphology,
NGC 3998 because of poor $R$ band calibration, and NGC 3896 
due to its poor $K'$ band data).  In this way, we have accumulated
a sample of 121 low-inclination 
spiral galaxies with radially-resolved, high-quality 
optical and near-IR data.

Radially-resolved colours from Paper {\sc i} and dJ{\sc iv}
were used in this study, 
and surface photometry from \tv96 was used to construct radially-resolved
colours.  Structural parameters and morphological types were taken 
from the sample's source papers and de Jong (1996a; dJ{\sc ii} hereafter). 
In the next two paragraphs we briefly outline the 
methods used to derive structural parameters and determine 
radially-resolved galaxy colours in the three source papers.

Galaxy structural parameters were determined using different
methods in each of the three source papers.  For the sample taken from 
dJ{\sc i},
a two-dimensional exponential bulge-bar-disc decomposition was used
\cite{djii}.  Paper {\sc i} 
uses a one-dimensional bulge-disc decomposition (with either 
an exponential or a r$^{1/4}$ law bulge).  
\tv96 use a `marking the disc' fit, where the contribution of the bulge
to the disc parameters is minimised by visually assessing where the
contribution from the bulge component is negligible.  
Note that dJ{\sc ii} shows that these methods all give 
comparable results to typically better than 20 per cent in terms of 
surface brightness and 10 per cent in terms of scale lengths, with little
significant bias.  
Magnitudes for all galaxies were determined from extrapolation of the
radial profiles with an exponential disc.
The dominant source of error in the determination of the galaxy
parameters is due to the uncertainty in sky level \cite{papi,djii}.

In order to examine radial trends in the stellar populations and galaxy
dust content, colours were determined in 
large independent radial bins to minimise the random errors in the
surface photometry and small scale stellar population and dust 
fluctuations.  Up to 7 radial bins (depending on the
signal-to-noise in the galaxy images) were used: 
$0 \le r/h_K < 0.5, 0.5 \le r/h_K < 1.5, 1.5 \le r/h_K < 2.5$ and so on,
where $r$ is the major-axis radius, and $h_K$ is
the major-axis $K$-band exponential disc scale length.  
The ellipticities and position angles of the annuli used to
perform the surface photometry were 
typically determined from
the outermost isophotes, and the centroid of the 
brightest portion of each galaxy 
was adopted as the ellipse centre.
Modulo calibration errors, the dominant source of uncertainty in the
surface photometry is due to the uncertainty in the adopted sky level
\cite{papi,djiv}.  

The galaxy photometry and colours were corrected for Galactic foreground
extinction using the dust models of Schlegel, Finkbeiner and
Davis \shortcite{sfd98}.  Ninety per cent of the galactic extinction
corrections in $B$ band are smaller than 0.36 mag, and 
the largest extinction correction is 1.4 mag.
We correct the $K$ band central surface brightness 
to the face-on orientation 
assuming that the galaxy is optically thin:  
$\mu^i_0  = \mu^{obs}_0 + 2.5 \log_{10}(\cos i)$, where 
$\mu^i_0$ is the inclination corrected surface brightness and
$\mu^{obs}_0$ is the observed surface brightness. 
The inclination $i$ is derived from 
the ellipticity of the galaxy $e$ assuming an intrinsic 
axial ratio $q_0$ of 0.15 \cite{holmberg1958}:  
$\cos^2 i = \{(1-e)^2-q_0^2\}/(1-q_0^2)$.
Assuming that the galaxies are optically thin in $K$ should
be a reasonable assumption as the extinction in $K$ is around ten 
times lower than the extinction in $B$ band.
Note that we do not correct the observed galaxy colours 
for the effects of dust:  the uncertainties that arise
from this lack of correction are addressed in section 
\ref{subsec:dust}.

The gas fraction of our sample is estimated as follows. 
The $K$ band absolute magnitude is converted into a stellar mass using
a constant stellar $K$ band mass to light ratio of 0.6 $M_{\sun}/L_{\sun}$
(c.f.\ Verheijen 1998; Chapter 6) and
a solar $K$ band absolute magnitude of 3.41 \cite{allen}.
The \hi mass is estimated using \hi fluxes from the NASA/IPAC
Extragalactic Database and the homogenised distances, increasing these masses
by a factor of 1.33 to account for the unobserved helium fraction
\cite{deblok1996}. 
We estimate molecular gas masses (also corrected for 
helium) using the ratio of molecular
to atomic gas masses as a function of morphological type 
taken from Young \& Knezek \shortcite{molec}.  The total gas mass is then
divided by the sum of the gas and stellar masses to form the gas fraction.
There are two main sources of systematic uncertainty in the
gas fractions not reflected in their error bars.
Firstly, stellar mass to light
ratios carry some uncertainty:  they
depend sensitively on the SFH and 
on the effects of dust extinction.  Our use of $K$ band to 
estimate the stellar masses to some extent alleviates these
uncertainties; however, it must be borne in mind that the stellar masses
may be quite uncertain.  Secondly, the molecular
gas mass is highly uncertain:  uncertainty in the CO to H$_2$ 
ratio, considerable scatter in the molecular to atomic gas mass ratio
within a given morphological type and the use of a type-dependent correction
(instead of e.g. a more physically motivated surface brightness or magnitude
dependent correction) all make the molecular gas mass a large uncertainty
in constructing the gas fraction.  This uncertainty could be better
addressed by using only galaxies with known CO fluxes; however, 
this requirement
would severely limit our sample size.  As our main aim is to search for
trends in age and metallicity (each of which is in itself quite uncertain)
as a function of galaxy parameters, a large sample size is more important
than a slightly more accurate gas fraction.
Note that this gas fraction is only an estimate of the 
cold gas content of the galaxy:  we do not consider any
gas in an extended, hot halo in this paper as it is unlikely
to directly participate in star formation, and therefore can be
neglected for our purposes.

\subsection{The stellar population models}

In order to place constraints on the stellar populations
in spiral galaxies, the 
galaxy colours must be compared with the colours of stellar
population synthesis (SPS) models.  
In this section, we outline
the SPS models that we use, and discuss the assumptions that we make
in this analysis.   We furthermore discuss the 
uncertainties involved in the use
of these assumptions and this colour-based technique.

In order to get some quantitative idea of the inaccuracies introduced
in our analysis by uncertainties in SPS
models, we used two different SPS models in our analysis:  
the {\sc gissel98} implementation of 
Bruzual \& Charlot (in preparation; hereafter BC98) and Kodama \& Arimoto
(1997; hereafter KA97).  
We use the multi-metallicity colour tracks of an
evolving single burst stellar population, with 
a Salpeter \shortcite{sp} stellar initial mass function (IMF)
for both models, where the lower mass 
limit of the IMF was 0.1 M$_{\odot}$ for both the
BC98 and KA97 models, and the upper mass limit was
125 M$_{\odot}$ for BC98 and 60 M$_{\odot}$ for the KA97 model. 
We assume that the IMF does not vary as a function of time.
We use simplified SFHs and fixed metallicities to explore
the feasible parameter space in matching the model colours with 
the galaxy colours.  Even though fixing the metallicities ignores chemical
evolution, the parameter space that we have used 
allows the determination of relative galaxy
ages and metallicities with a minimal set of assumptions. 

In this simple case, the integrated spectrum $F_{\lambda}(t)$ for a
stellar population with an arbitrary star formation rate (SFR) $\Psi(t)$ is
easily obtained from the time-evolving spectrum of a single-burst
stellar population with a given metallicity 
$f_{\lambda}(t)$ using the convolution 
integral \cite{bc93}:
\begin{equation}
F_{\lambda}(t) = \int_0^t \Psi(t-t') f_{\lambda}(t') dt'.
\end{equation}
We use exponential SFHs,
parameterised by the star formation timescale $\tau$.  In this
scenario, the SFR $\Psi(t)$ is given by:
\begin{equation}
\Psi(t) = B e^{-t/\tau},  \label{eqn:sfr}
\end{equation} 
where $B$ is an arbitrary constant determining the total mass of the
stellar population.  In order to cover the entire range of colours in
our sample, both exponentially decreasing and increasing SFHs
must be considered.  Therefore, $\tau$ is an
inappropriate choice of parameterisation for the SFR,
as in going from old to young stellar populations smoothly, 
$\tau$ progresses from 0 to $\infty$ for a constant SFR,
then from $-\infty$ to small negative values.
We parameterise the SFH by 
the average age of the stellar population $\langle A \rangle$, given by:
\begin{equation}
\langle A \rangle = \frac{\int_0^A t \Psi(t) dt}{\int_0^A \Psi(t) dt} = 
      A - \tau \frac{1 - e^{-A/\tau}(1 + A/\tau)}{1 - e^{-A/\tau}}
\end{equation}
for an exponential SFH, as given in Equation
\ref{eqn:sfr}, where $A$ is the age of the oldest stars in the stellar
population (signifying when star formation started).  In our case, we
take $A = 12$ Gyr as the age of all galaxies, and parameterise the
difference between stellar populations purely in terms of $\langle A \rangle$.

Clearly, the above assumptions that we use to construct a grid of model
colours are rather simplistic.  In particular, our assumption
of a galaxy age of 12 Gyr and an exponential SFH,
while allowing almost complete coverage of the colours observed in our sample,
is unlikely to accurately reflect the broad range of 
SFHs in our diverse sample of galaxies.  Some
galaxies may be older or younger than 12 Gyr, and star
formation is very likely to proceed in bursts, instead of
varying smoothly as we assume above.  Additional uncertainties
stem from the use of a constant stellar metallicity and from 
the uncertainties inherent to the SPS models themselves, which 
will be at least $\sim$ 0.05 mag for the optical passbands, increasing 
to $\ga$ 0.1 mag for the near-IR passbands \cite{charlot1996}.

However, the important point here is that the above method gives
robust {\it relative} ages and metallicities.  
Essentially, this colour-based method gives some kind of 
constraint on the luminosity-weighted 
ratio of $\la$ 2 Gyr old 
stellar populations to $\ga$ 5 Gyr old stellar populations, which 
we parameterise using the average age $\langle A \rangle$.
Therefore, the details of how we construct the older and younger 
stellar populations are reasonably unimportant.  It is perfectly possible
to assume an exponential SFH with an age of only 5 Gyr:
if this is done, galaxies become younger (of course), but the relative
ordering of galaxies by e.g.\ age is unaffected.
Note that because the colours are luminosity weighted, small
bursts of star formation may affect the relative 
ages and metallicities of galaxies (relative to their underlying ages and 
metallicities before the burst); however, it is unlikely that a 
large fraction of galaxies will be strongly affected by large amounts
of recent star formation.
The basic message is that absolute galaxy ages and metallicities
are reasonably uncertain, but the relative trends are robust.
Note that all of the results presented in this paper were
obtained using the BC98 models, unless otherwise stated.

\subsection{The dust models} \label{subsec:dustmod}

We {\it do not} use dust models in the SFH fitting; however, we use 
dust models in Figs.\ \ref{fig:colcoltype}--\ref{fig:colcolfg} and
section \ref{subsec:dust} to allow us to quantify the effects 
that dust reddening may have on our results.  
We adopt the Milky Way (MW) and Small Magellanic Cloud (SMC) 
extinction curves and albedos from Gordon, Calzetti \& Witt 
\shortcite{gordon1997}.  In the top two panels in Figs.\ 
\ref{fig:colcoltype}--\ref{fig:colcolfg}, we show a 
MW foreground screen model with a total $V$ band 
extinction $A_V = 0.3$ mag (primarily to facilitate
comparison with other studies).  In the lower two panels, 
we show the reddening vector from a more realistic face-on
Triplex dust model 
\cite[Disney, Davies \& Phillips 1989; DDP hereafter]{evans1994}.  
In this model, the dust and stars are distributed smoothly 
in a vertically and horizontally
exponential profile, with equal vertical and horizontal distributions.  
The two models shown have a reasonable central $V$ band optical depth
in extinction, from pole to pole, of two.
This value for the central optical depth in $V$ band is 
supported by a number of recent statistical studies into the 
global properties of dust in galaxies 
\cite{pelt92,hu94,tv97,tully1998,k98}.
However, this model does not
take account of scattering.  
Monte Carlo simulations of Triplex-style galaxy models
taking into account the effects of scattering indicate
that because our sample is predominantly face-on, 
at least as many photons will be scattered into the line of sight as 
are scattered out of the line of sight \cite{djiv}.  Therefore, 
we use the dust {\it absorption} curve to approximate
the effects of more realistic distributions of dust on the 
colours of a galaxy.  The use of the absorption curve is the main reason for
the large difference in the MW and SMC Triplex model vectors: 
Gordon et al.'s \shortcite{gordon1997} 
MW dust albedo is much higher than their SMC dust albedo, 
leading to less absorption per unit pole to pole extinction.  
Note also that the Triplex dust reddening vector produces much more 
optical reddening than e.g.\ those in dJ{\sc iv}:  this is due to 
our use of more recent near-IR dust albedos, which are much larger
than e.g.\ those of Bruzual, Magris \& Calvet \shortcite{bruzual1988}. 

\section{Maximum-likelihood fitting} \label{sec:max}

\begin{figure}
\begin{center}
  \leavevmode
  \epsffile{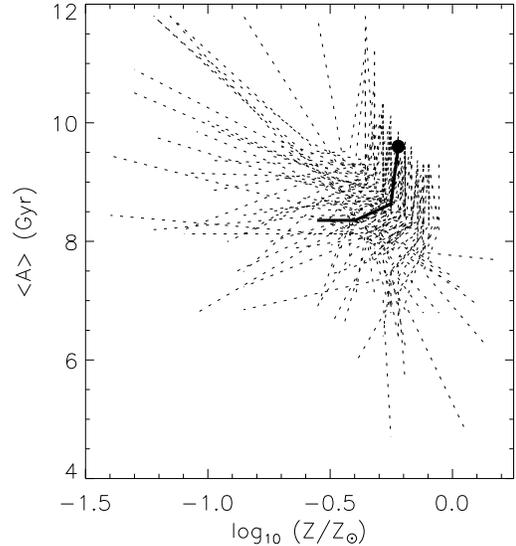}
\end{center}
\caption{An example of the derived ages and metallicities, with their  
  associated Monte Carlo error estimates, for UGC 3080.  The solid line 
  denotes the best fit age and metallicities of UGC 3080
  as a function of radius, in bins of $0 \le r/h_K < 0.5$ (filled circle), 
  $0.5 \le r/h_K < 1.5, 1.5 \le r/h_K < 2.5$ and  
  $2.5 \le r/h_K < 3.5$.  The dotted lines denote the 100
  Monte Carlo simulations, with
  realistic photometric errors applied.  The galaxy centre seems both
  reasonably old and metal rich, and becomes younger and more
  metal poor as you go further out into the spiral disc.  Note that the
  Monte Carlo simulation indicates that these trends are reasonably significant
  for this galaxy.  
}
\label{fig:u3080}
\end{figure}

Before we describe the maximum-likelihood fitting method in detail, it
is important to understand the nature of the error budget in an
individual galaxy colour profile.  There are three main contributions
to the galaxy photometric errors {\it in each passband} 
(note that because of the use of large radial bins, shot noise errors are
negligible compared to these three sources of error).
\begin{enumerate}
\item Zero-point calibration errors $\delta\mu_{cal}$ 
  affect the whole galaxy colour profile systematically by
  applying a constant offset to the colour profiles.
\item In the optical passbands, flat fielding errors $\delta\mu_{ff}$ 
  affect the galaxy colour profile randomly as
  a function of radius.  Because of the use of large
  radial bins in the construction of the colours, the dominant
  contribution to the flat fielding uncertainty is the flat field
  structure over the largest scales.  Because the
  flat fielding structure is primarily large scale, it is fair
  to reduce the flat fielding error by a factor of $\sim$ 10 as 
  one radial bin will only cover typically one tenth or less 
  of the whole frame. 
  We assume normally-distributed flat fielding
  errors with a $\sigma$ of 0.5/10
  per cent of the galaxy and sky level in the optical (0.5 per cent
  was chosen as a representative, if not slightly generous, flat field
  error over large scales).
  In the near-IR, sky subtraction to a large extent 
  cancels out any flat fielding uncertainties:  we, therefore, 
  do not include the effect of flat fielding uncertainties in
  the error budget for near-IR passbands.
\item Sky level estimation errors $\delta\mu_{sky}$
  affect the shape of the colour
  profile in a systematic way.  If the assumed sky level is too high,
  the galaxy profile `cuts off' too early, making the galaxy appear
  smaller and fainter than it is, whereas if the sky level is
  too low the galaxy profile `flattens off' and the galaxy appears
  larger and brighter than it actually is.  As long as the final
  estimated errors in a given annulus due to sky variations are
  reasonably small ($\la 0.3$ mag)
  this error $\delta\mu_{sky}$ is related to the sky level error $\delta\mu_s$ 
  (in magnitudes) by:
  \begin{equation}
    \delta\mu_{sky} \simeq \delta\mu_s 10^{0.4(\mu - \mu_s)},
  \end{equation}
  where $\mu$ is the average surface brightness of the galaxy in 
  that annulus and $\mu_s$ is the sky surface brightness.
\end{enumerate}
These photometric errors have 
different contributions to the total error budget at different radii, and have
different effects on the ages, metallicities, and their gradients.
If a
single annulus is treated on its own, the total error estimate for each
passband $\delta\mu_{tot}$ is given by:
\begin{equation}
  \delta\mu_{tot} = 
     (\delta\mu_{cal}^2 + \delta\mu_{ff}^2 + \delta\mu_{sky}^2)^{1/2}.
\end{equation}
We use magnitude errors instead of flux errors in 
constructing these error estimates:  this simplifies the 
analysis greatly, and is more than accurate enough for 
our purposes.

We fit the SPS models to each set of radially-resolved galaxy colours using a
maximum-likelihood approach.  Note that we do {\it not} include the 
effects of dust in the maximum-likelihood fit.
\begin{itemize}  
\item We generate a finely-spaced model grid by
calculating the SPS models for a fine grid of $\tau$ values.  These
finely spaced $\tau$ grids are then interpolated along the metallicity 
using a cubic spline between the model metallicities.
These models then provide a set of arbitrary normalisation model magnitudes
$\mu_{{\rm model},i}(\langle A \rangle,Z)$ for each passband $i$ 
for a range of SFHs and metallicities.  
\item We read in the multi-colour 
galaxy surface brightnesses as a function of radius, {\it treating
each annulus independently} (each annulus has surface brightness 
$\mu_{{\rm obs},i}$).
The total error estimate
$\delta\mu_{tot,i}$ for each annulus 
is determined for each passband $i$.  
\item For each $\langle A \rangle$ 
and metallicity $Z$, it is then possible to determine the best
normalisation between the model and annulus colours $\mu_c$ using: 
\begin{equation}
\mu_c = \frac{\sum_{i=1}^n \{\mu_{{\rm obs},i} - 
	\mu_{{\rm model},i}(\langle A \rangle,Z)\}/\delta\mu_{tot,i}^2}
      {\sum_{i=1}^n 1/\delta\mu_{tot,i}^2},
\end{equation} 
where $n$ is the number of passbands 
and its corresponding $\chi^2$ is given by:
\begin{equation}
\chi^2 = \frac{1}{n-1} \sum_{i=1}^n \frac{(\mu_{{\rm obs},i} - 
	\mu_{{\rm model},i}(\langle A \rangle,Z) - \mu_c)^2}
         {\delta\mu_{tot,i}^2}.
\end{equation}
The best model $(\langle A \rangle,Z)$ match is the one with the
minimum $\chi^2$ value.  This procedure is then carried out for all of
the galaxy annuli.  
Note that minimising $\chi^2$, strictly speaking,
requires the errors to be Gaussian; however, using magnitude errors
$\la$ 0.3 mag does not lead to significant errors in $\chi^2$.
\end{itemize}

\subsection{Estimating age and metallicity gradients} \label{sec:grad}

\begin{figure*}
\begin{minipage}{175mm}
\begin{center}
  \leavevmode
  \epsffile{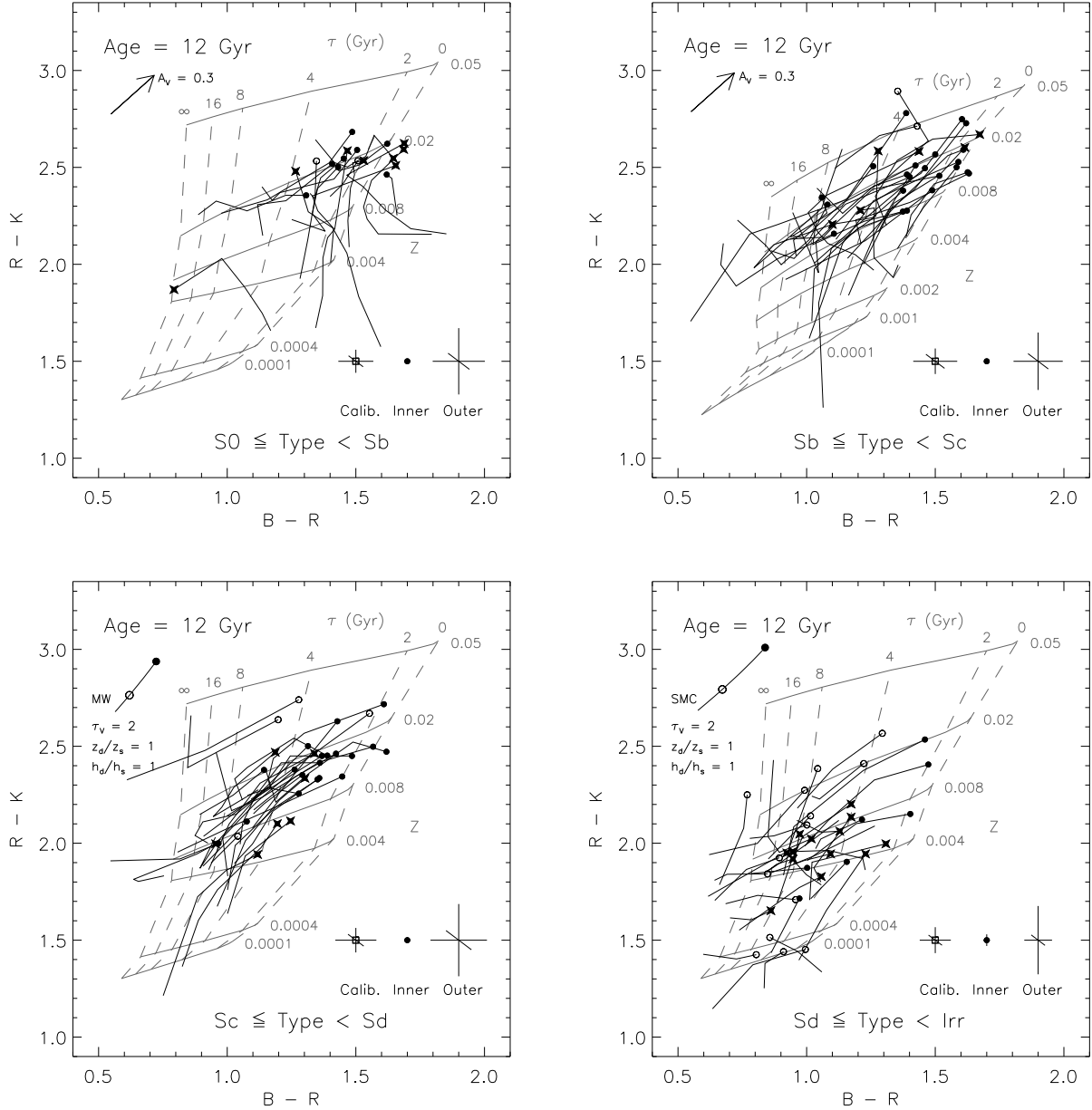}
\end{center}
\caption{  
Trends in optical--near-IR colour with morphological type.
We plot $B - R$ against $R - K$ colours for those 
galaxies with photometry in $B$, $R$ and $K$ 
(92 per cent of the sample; solid lines).  Central colours are
denoted by solid circles (for the sample from dJ{\sc iv}), 
open circles (for the sample from Paper {\sc i}) or 
stars (for the sample from TVPHW).  Overplotted are the
KA97 (upper right panel) and BC98 (remaining panels) 
stellar population models.  
Note that the lower right panel is labelled with the average 
age of the stellar population $\langle A \rangle$; the remaining
panels are labelled with the star formation timescale $\tau$.
We also show the calibration uncertainties
and the sky subtraction uncertainties in the inner and outer
annuli:  the diagonal lines in these error bars represent
the effect of a 1$\sigma$ shift in the $R$ band photometry.  
Overplotted also are the foreground screen (upper panels)
and Triplex model (lower panels) dust models.  
In the Triplex model, the solid circle denotes the central reddening,
and the open circle the reddening at the half light radius.
Further discussion of
the data and model grids is undertaken in the text.
}
\label{fig:colcoltype}
\end{minipage}
\end{figure*}

\begin{figure*}
\begin{minipage}{175mm}
\begin{center}
  \leavevmode
  \epsffile{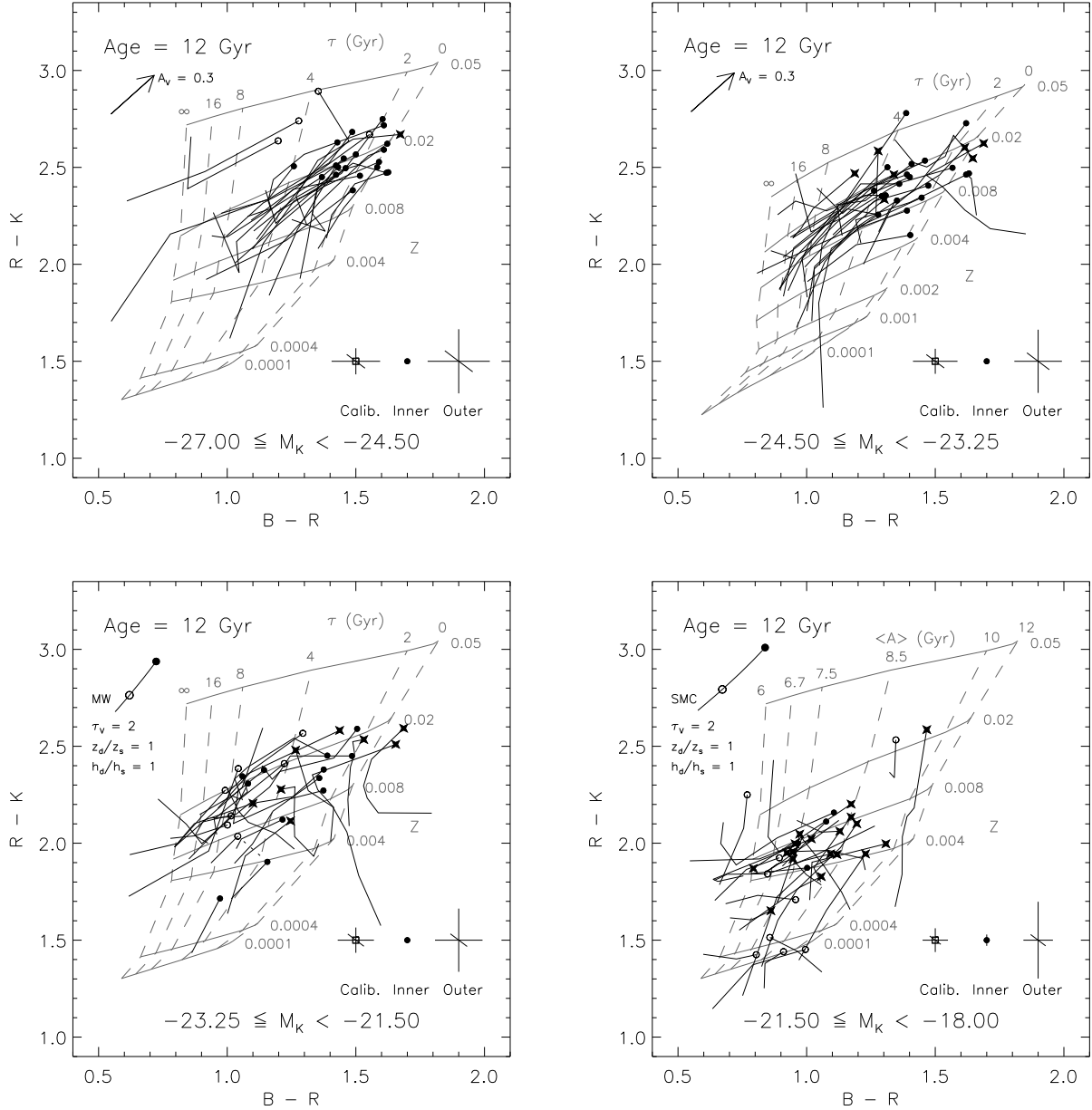}
\end{center}
\caption{  
Trends in optical--near-IR colour with $K$ band magnitude.
The symbols are as in Fig.\,\protect\ref{fig:colcoltype}.
The two sets of colours for UGC 334 from dJ{\sc iv} and Paper {\sc i} 
are also shown, connected by a dotted line.}
\label{fig:colcolmag}
\end{minipage}
\end{figure*}

\begin{figure*}
\begin{minipage}{175mm}
\begin{center}
  \leavevmode
  \epsffile{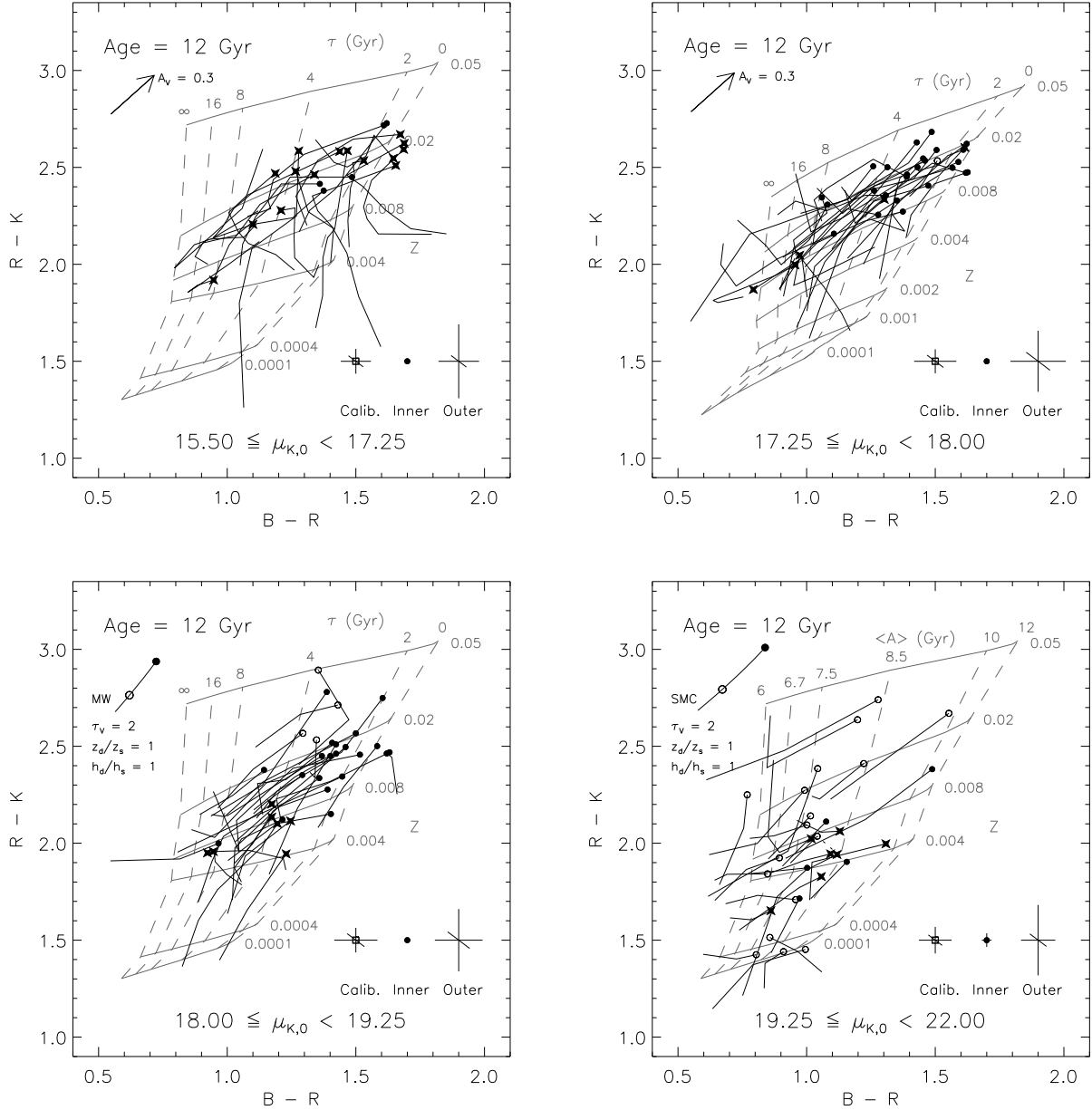}
\end{center}
\caption{  
Trends in optical--near-IR colour with $K$ band central surface brightness.
The symbols are as in Fig.\,\protect\ref{fig:colcoltype}.
The two sets of colours for UGC 334 from dJ{\sc iv} and Paper {\sc i} 
are also shown, connected by a dotted line.}
\label{fig:colcolmu}
\end{minipage}
\end{figure*}

\begin{figure*}
\begin{minipage}{175mm}
\begin{center}
  \leavevmode
  \epsffile{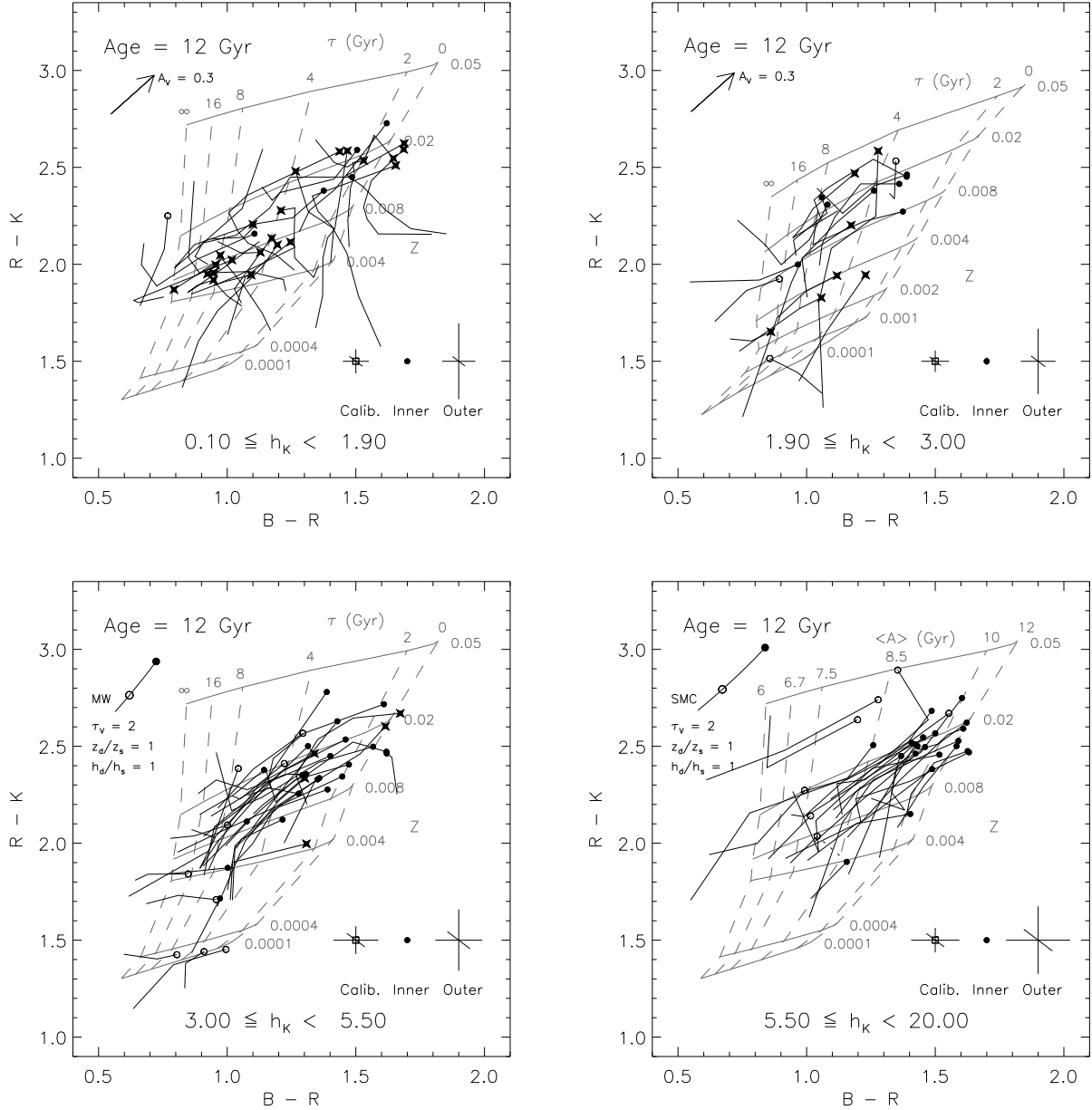}
\end{center}
\caption{  
Trends in optical--near-IR colour with $K$ band disc scale length in kpc.
The symbols are as in Fig.\,\protect\ref{fig:colcoltype}.
The two sets of colours for UGC 334 from dJ{\sc iv} and Paper {\sc i} 
are also shown, connected by a dotted line.}
\label{fig:colcolscal}
\end{minipage}
\end{figure*}

\begin{figure*}
\begin{minipage}{175mm}
\begin{center}
  \leavevmode
  \epsffile{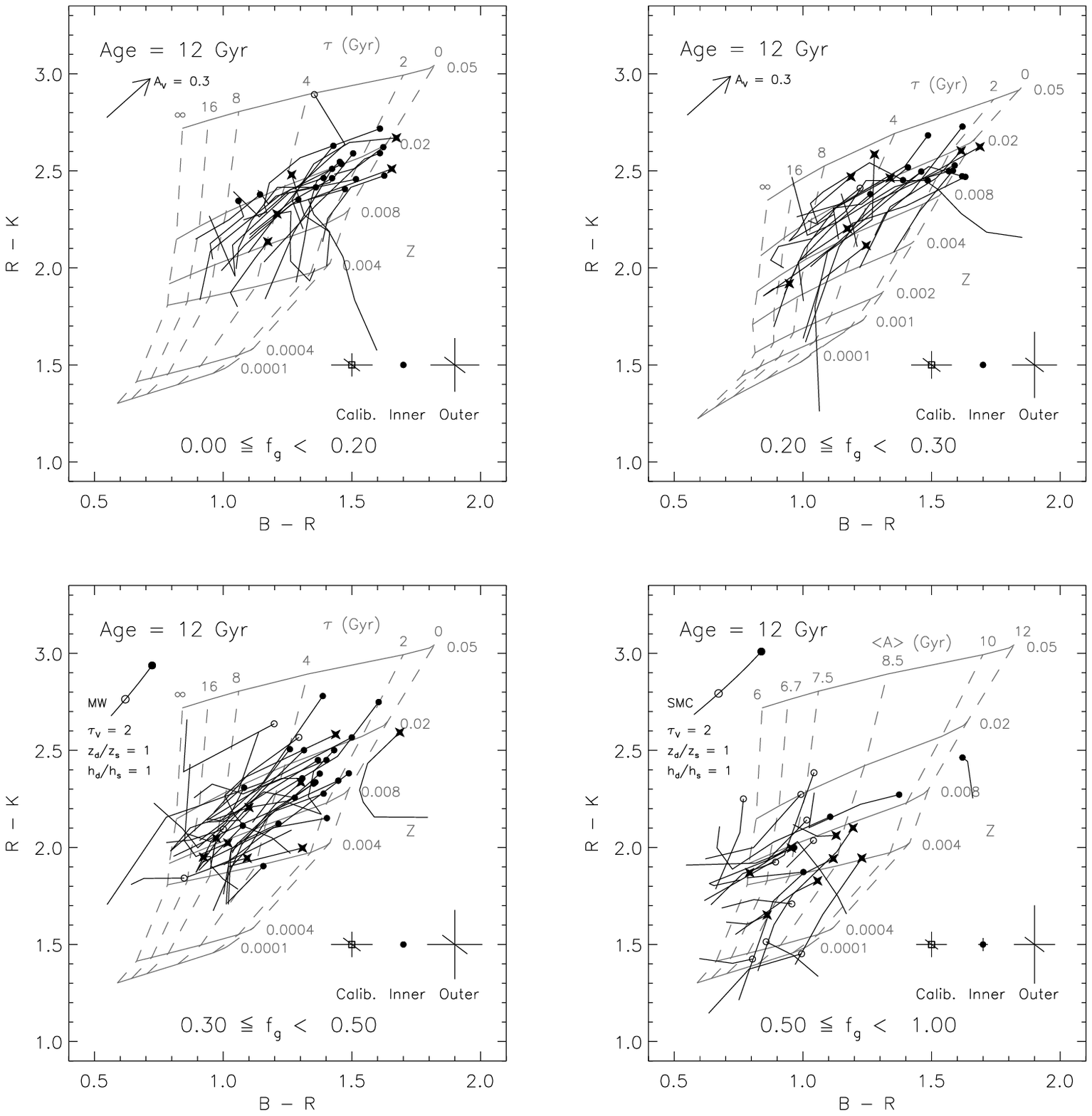}
\end{center}
\caption{  
Trends in optical--near-IR colour with gas fraction.
The symbols are as in Fig.\,\protect\ref{fig:colcoltype}.}
\label{fig:colcolfg}
\end{minipage}
\end{figure*}

In order to gain insight into age and metallicity trends as a function 
of radius, we perform a {\it weighted} linear fit to the ages and
metallicities of each galaxy.  This fit is parameterised 
by a gradient per disc scale length and an
intercept at the disc half light radius.  

Each point entering the fit is weighted by the total 
age or metallicity error in each annulus, determined using
$\Delta\chi^2 = 1$ intervals for the age and metallicity parameters
individually (i.e.\ the age or metallicity
range for which the reduced $\chi^2$ is smaller than 
$\chi_{\rm best\,fit}^2 + 1$).  
The first two, three or four bins of
the radially-resolved ages and metallicities are used in the fit.
Galaxies with only central colours are omitted from the sample,
and galaxies with more than four bins are fit only out to the fourth bin.
Including the fifth, sixth and seventh bins in the
gradient fits does not significantly
affect our results, but can increase the noise in the gradient
determinations.

As two out of the three sources of photometric error are correlated
over the galaxy, it is impossible to use the $\Delta\chi^2 = 1$ contours
for each annulus to estimate e.g.\ the errors in age and metallicity
gradients.  This is because to a large extent
calibration uncertainties will not affect the 
age or metallicity gradient; however, the errors for the age or
metallicity intercept will depend sensitively 
on the size of calibration error.
We, therefore, produce error estimates using a Monte Carlo approach.

For each galaxy, we repeat the above SFH
derivation and analysis 100 times,
applying a normally distributed random calibration, flat fielding
and sky level errors.
This approach has the virtue that the different types of
photometric error act on the galaxy profile as they should, allowing
valid assessment of the accuracy of any trends in age or metallicity
with radius.  An example of the ages and metallicities
derived with this approach for the galaxy UGC 3080 is presented
in Fig. \ref{fig:u3080}.  The solid line indicates the best-fit model
for the galaxy colours as a function of radius, and the dotted lines
the Monte Carlo simulations with realistic (or 
perhaps slightly conservative) photometric errors.  Note
that the errors in age and metallicity increase substantially 
with radius.  This is
primarily due to the sky subtraction errors, as their effect increases
as the surface brightness becomes fainter with radius.  Note  
that because we use a weighted fit, 
the age and metallicity gradients are robust to the largest of 
the sky subtraction errors in the outer parts.

The error in any one parameter, e.g.\ metallicity, in any one 
annulus is given by the half of the interval containing 
68 per cent of the e.g.\ metallicities in that annulus (the 
one sigma confidence interval).
Errors in the age or metallicity gradients and intercepts are determined by
fitting the 100 Monte Carlo simulated datasets in the same way as the
real data.  As above, the errors in each of these 
fit parameters are determined by
working out the one sigma confidence interval of the simulations 
for each fit parameter alone.

\section{Results} \label{sec:res}

\subsection{Colour-colour plots} \label{sec:colcol}

In order to better appreciate the characteristics of the data,
the model uncertainties and the potential 
effects of dust reddening on the results, it is useful to consider some
colour-colour plots in some detail.  In Figs.\
\ref{fig:colcoltype} through 
\ref{fig:colcolfg} we
show $B - R$ against $R - K$ colours  as a function of radius for our sample
galaxies.  Central colours are
denoted by solid circles (for the sample from dJ{\sc iv}), 
open circles (for the sample from Paper {\sc i}) or 
stars (for the sample from TVPHW).    
Overplotted is the SPS model of KA97 in the
upper right panel; in the remaining panels we plot BC98's model.  
Note that the lower right panel is labelled with the average 
age of the stellar population $\langle A \rangle$; the remaining
panels are labelled with the star formation timescale $\tau$.
In the upper panels we show
a dust reddening vector assuming that dust is distributed in a
foreground screen, and in the lower panels we show two absorption reddening
vectors from the analytical treatment of DDP and Evans \shortcite{evans1994}. 

These figures clearly reiterate the conclusions of Paper {\sc i}
and dJ{\sc iv}, where it was found that colour gradients were common in
all types of spiral discs.  Furthermore, these colour
gradients are, for the most part, consistent with 
gradients in mean stellar age.  The effects of dust reddening
may also play a r\^{o}le; however, dust is unlikely 
to produce such a large colour gradient:  see section \ref{subsec:dust}
and e.g.\ Kuchinski et al.\ \shortcite{k98} for a more detailed
justification.  

A notable exception to this
trend are the relatively faint S0 galaxies from the sample of TVPHW 
(see Fig.\,\ref{fig:colcoltype}).  These galaxies
show a strong metallicity gradient, and an inverse age gradient, 
in that {\it the outer regions of the galaxy appear older and more 
metal poor} than the younger and more metal rich central regions of
the galaxy.  This conclusion is consistent with the results from 
studies of line strength gradients:  Kuntschner \shortcite{haraldthesis}
finds that many of the fainter S0s in his sample have a relatively
young and metal rich nucleus, with older and more metal poor outer regions.

Also, in Fig. \ref{fig:colcoltype}, we find that there 
is a type dependence in the galaxy age and metallicity, in the sense that
later type galaxies are both younger and more metal poor than
their earlier type counterparts.
This is partially consistent with 
dJ{\sc iv}, who finds that later type spirals have predominantly
lower stellar metallicity (but he finds no significant trends in
mean age with galaxy type).

In Figs. \ref{fig:colcolmag} to \ref{fig:colcolfg} we explore the
differences in SFH as a function of the 
physical parameters of our sample galaxies.  
Figure \ref{fig:colcolscal} suggests there is little correlation
between galaxy scale length and SFH:  
there may be a weak correlation in the sense that 
there are very few young and metal poor large scale length
galaxies.
In contrast, when taken together, 
Figs. \ref{fig:colcolmag}, \ref{fig:colcolmu} and \ref{fig:colcolfg}
suggest that there are real correlations between 
the SFHs of spiral galaxies (as probed by 
the ages and metallicities) and the magnitudes, central
surface brightnesses and gas fractions of galaxies in 
the sense that brighter, higher surface brightness galaxies 
with lower gas fractions tend to be older and more metal rich than
fainter, lower surface brightness galaxies with larger gas fractions.
Later, we will see these trends more clearly in terms of 
average ages and metallicities, but it is useful to bear in 
mind these colour-colour plots when considering the trends 
presented in the remainder of this section.

\subsection{Local ages and metallicities}


\begin{figure*}
\begin{minipage}{175mm}
\begin{center}
  \leavevmode
  \epsffile{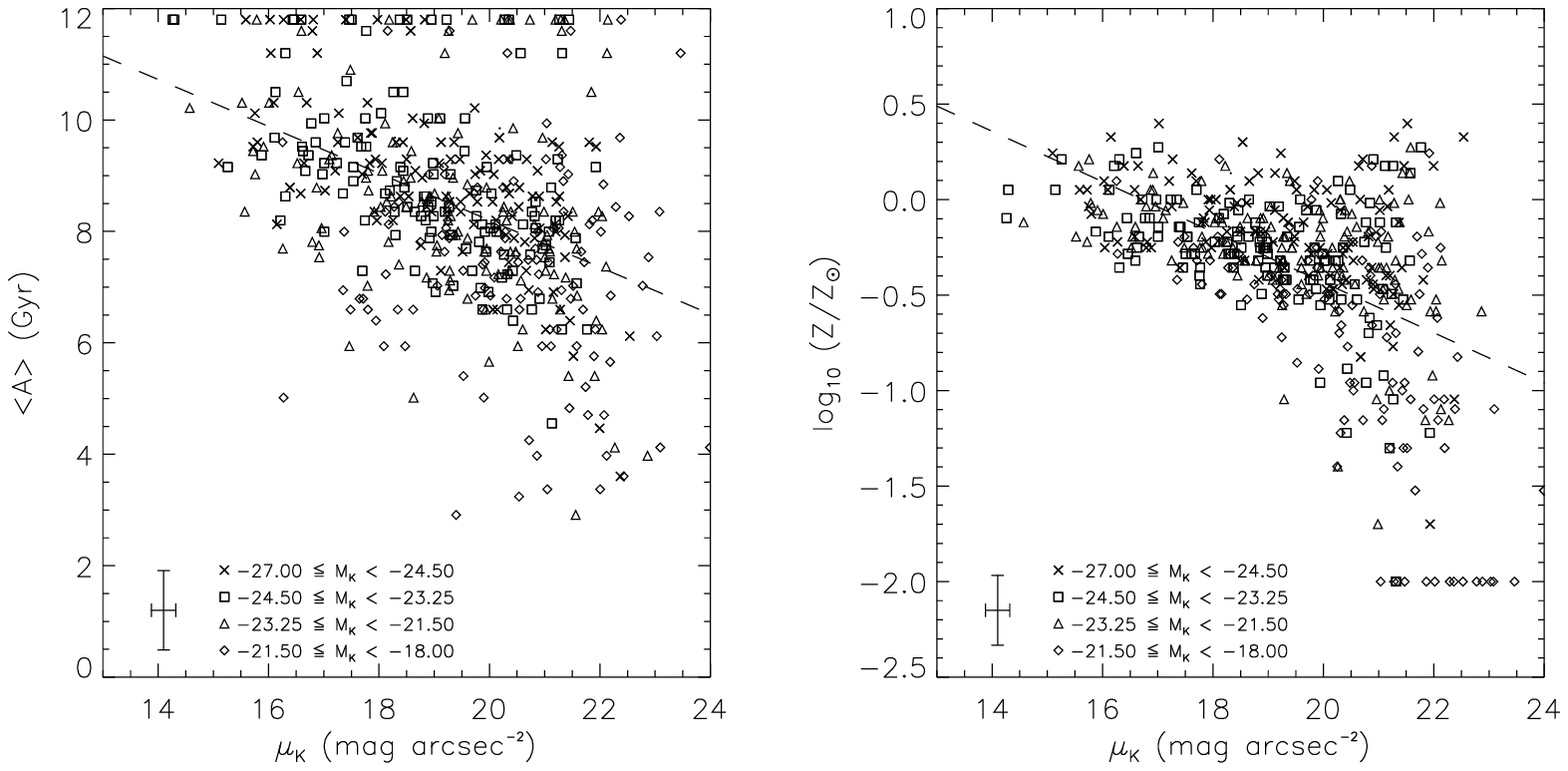}
\end{center}
\caption{  
Local average age (left) and metallicity (right) against $K$ band
surface brightness.  The average age $\langle A \rangle$ and
metallicity $\log_{10}(Z/Z_{\sun})$ of each 
galaxy annulus is shown against the local
average $K$ band surface brightness $\mu_K$. 
The galaxies are binned by $K$ band absolute magnitude:  note that
this binning clearly shows the magnitude dependence in the 
local metallicity--surface brightness correlation. 
The dashed lines are unweighted least-squares fits to the data:  the fits and 
significances are given in Table \protect\ref{tab:fit}.  
A significant number of data points have average ages of 11.9 Gyr or
metallicities of $\log_{10}(Z/Z_{\sun}) = -2$: these
data points fall outside the model grids.
}
\label{fig:local}
\end{minipage}
\end{figure*}


We investigate the relation between the average ages and metallicities inferred
from the colours of each galaxy annulus and the average $K$ band surface
brightness in that annulus in Fig. \ref{fig:local}.  
Representative error bars from the Monte Carlo simulations are
shown in the lower left hand corner of the plots.
Strong, statistically significant
correlations between local age and the $K$ band surface brightness, 
and between local metallicity and the $K$ band surface brightness are found.  
While the scatter
is considerable, the existence of such a strong relationship between
local $K$ band 
surface brightness and the age and metallicity in that region is remarkable.  
In order to probe the dependence of the SFH on the
structural parameters of our sample, we have carried
out {\it unweighted} least-squares fits of 
these and subsequent correlations: 
the results of these fits are outlined in Table \ref{tab:fit}.
In Table \ref{tab:fit} we give the coefficients of the unweighted
least-squares fits to the correlations
significant at the 99 per cent level (where the 
significances are determined from a Spearman rank order test), 
along with errors
derived from bootstrap resampling of the data \cite{numrec}.
We will address these
correlations further in the next section and in the discussion (section
\ref{sec:disc}), where we attempt to relate
these strong local correlations with global correlations.

\subsection{Global relations} \label{subsec:global}


\begin{figure*}
\begin{minipage}{175mm}
\begin{center}
  \leavevmode
  \epsffile{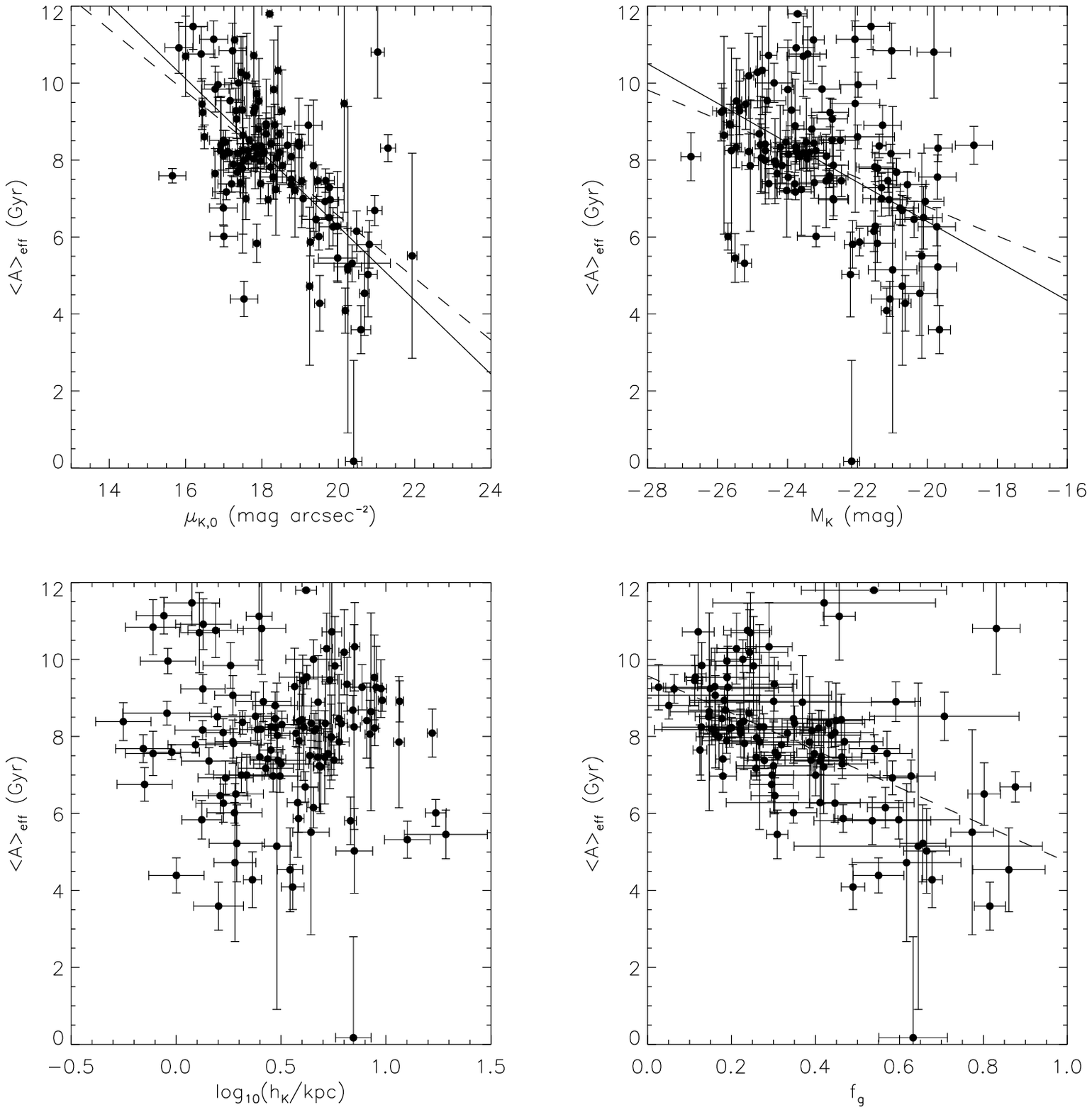}
\end{center}
\caption{
Correlations between the average age 
$\langle A \rangle_{eff}$
at the disc half
light radius and the $K$ band central surface brightness $\mu_{K,0}$,
$K$ band absolute magnitude $M_K$, $K$ band disc scale length $h_K$
and gas fraction $f_g$ (see text for the details of the 
derivation of the gas fractions).
In this and subsequent figures error bars
denote the estimated 68 per cent confidence intervals, 
and dashed lines are unweighted least-squares fits to trends significant
at greater than the 99 per cent level (see Table \protect\ref{tab:fit}).
The solid lines in the top two panels denote the unweighted least-squares
fit to the age--magnitude--surface brightness correlation, projected
onto the age--surface brightness and age--magnitude planes.
}
\label{fig:age}
\end{minipage}
\end{figure*}

\begin{figure*}
\begin{minipage}{175mm}
\begin{center}
  \leavevmode
  \epsffile{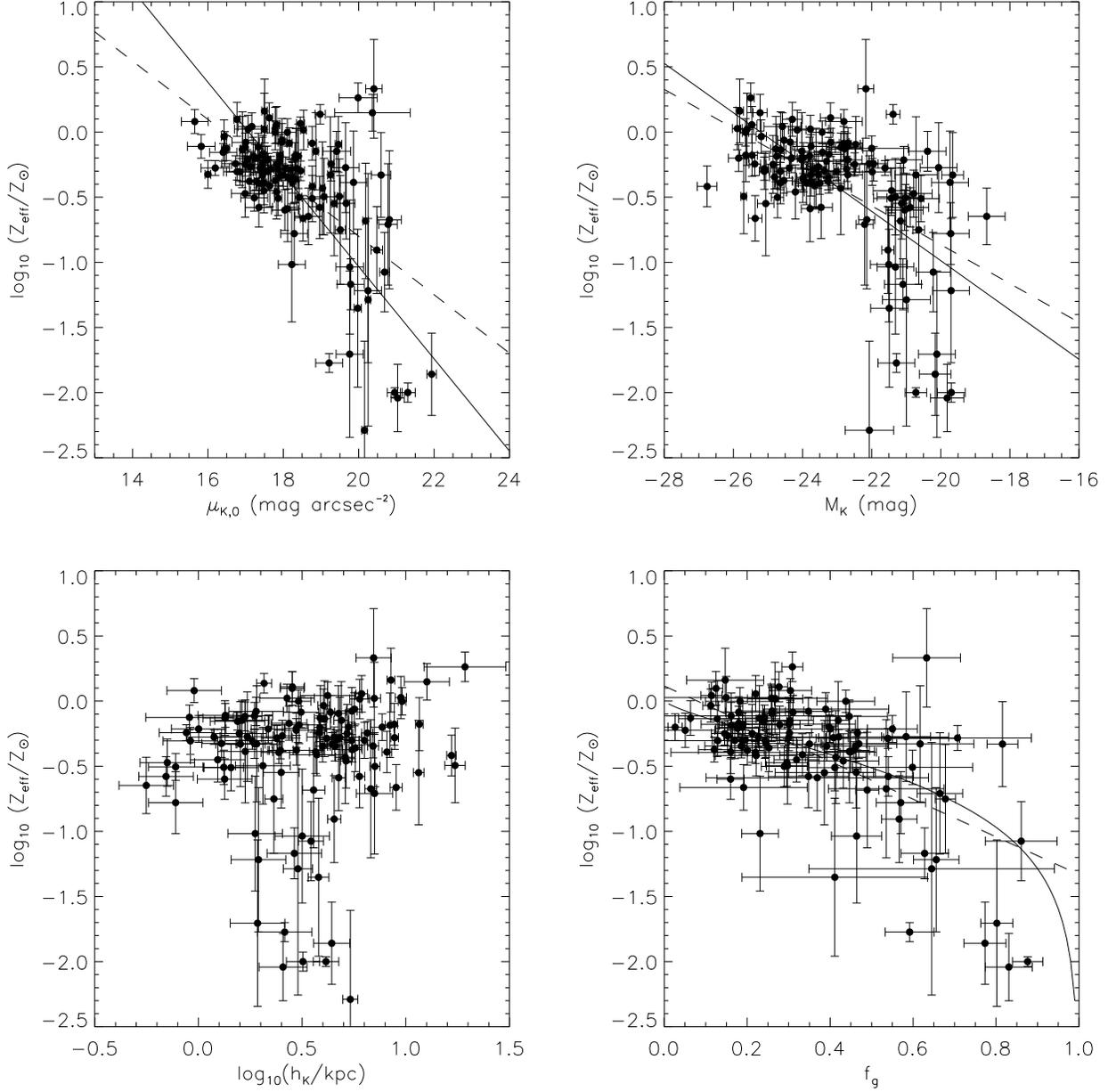}
\end{center}
\caption{  
Correlations between the average 
metallicity $\log_{10}(Z_{eff}/Z_{\sun})$ at the disc half
light radius and $K$ band central surface brightness, 
$K$ band absolute magnitude, $K$ band disc scale length and gas fraction.
Dashed lines are unweighted least-squares fits to trends significant
at greater than the 99 per cent level (see Table \protect\ref{tab:fit}).
The solid lines in the top two panels denote the unweighted least-squares
fit to the metallicity--magnitude--surface brightness correlation, projected
onto the metallicity--surface brightness and metallicity--magnitude planes.
The solid line in the lower right panel shows the expectation for the 
mean stellar metallicity with gas fraction for a closed-box model
with a solar metallicity yield.
}
\label{fig:met}
\end{minipage}
\end{figure*}

\begin{figure*}
\begin{minipage}{175mm}
\begin{center}
  \leavevmode
  \epsffile{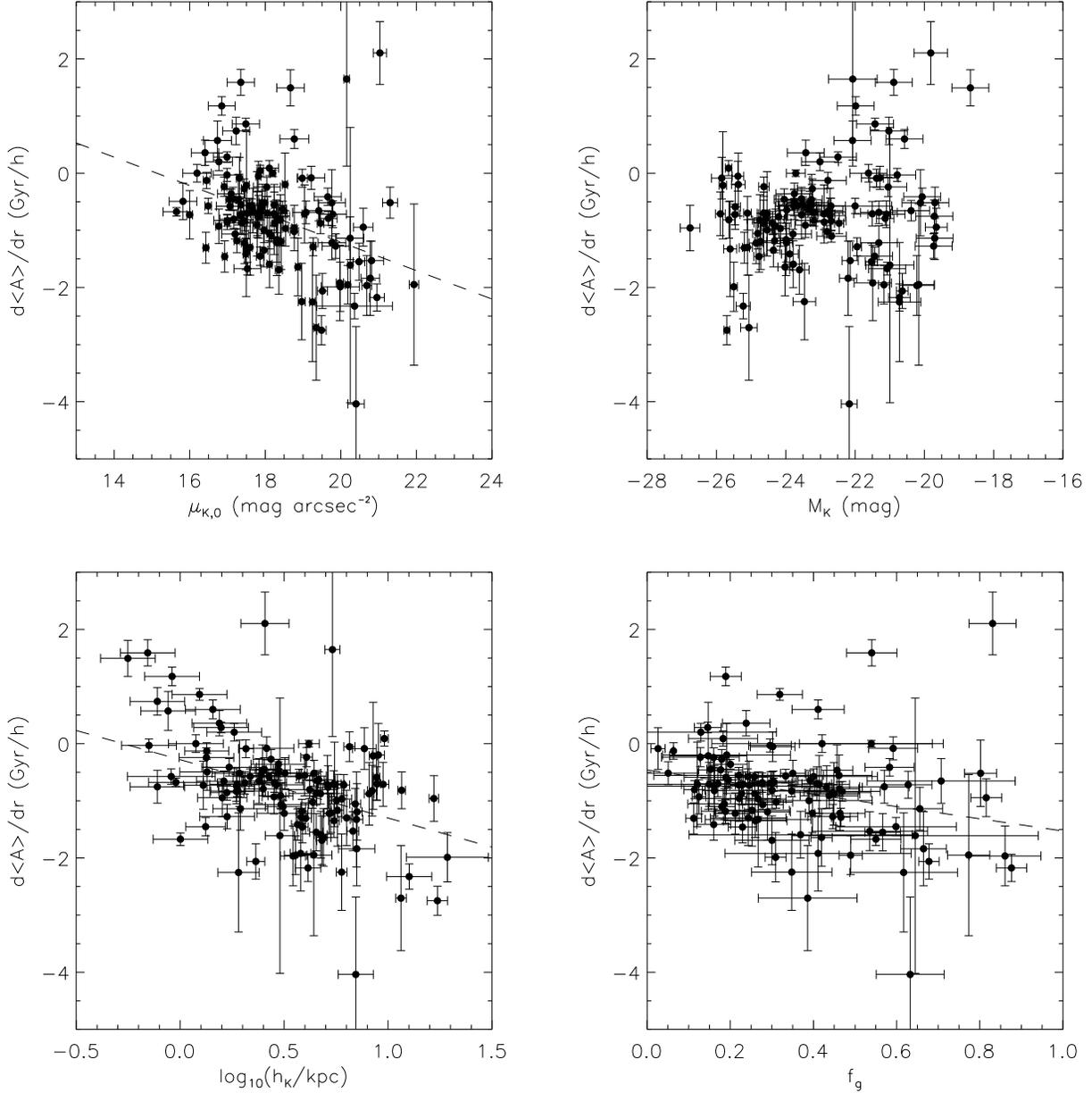}
\end{center}
\caption{
Correlations between the average age gradient 
per $K$ band disc scale length 
($d\langle A \rangle/dr$)
and the $K$ band central surface brightness, 
$K$ band absolute magnitude, $K$ band disc scale length and the gas fraction.
Dashed lines are unweighted least-squares fits to trends significant
at greater than the 99 per cent level (see Table \protect\ref{tab:fit}).
}
\label{fig:agegrad}
\end{minipage}
\end{figure*}

\begin{figure*}
\begin{minipage}{175mm}
\begin{center}
  \leavevmode
  \epsffile{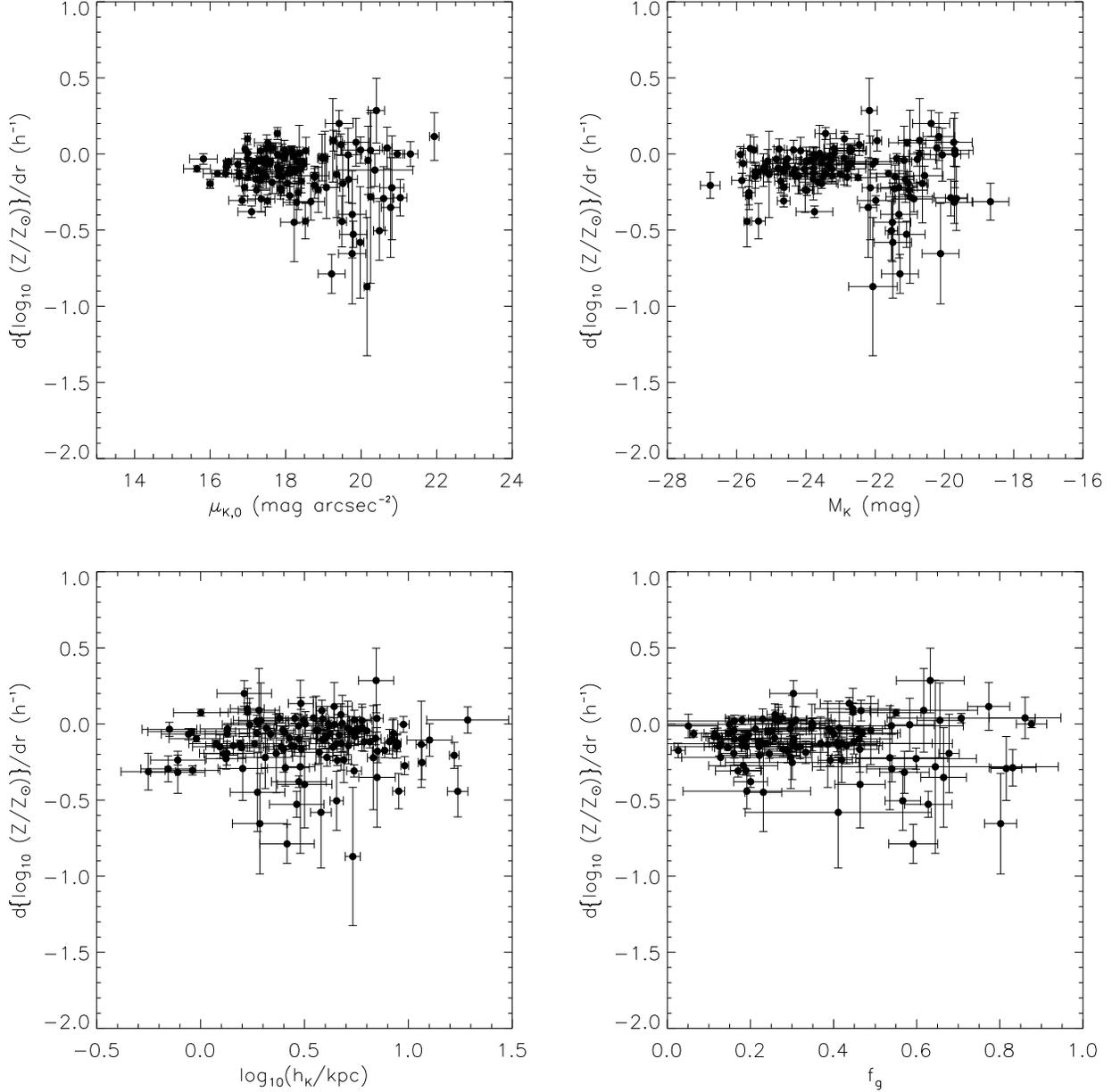}
\end{center}
\caption{  
Correlations between the average 
metallicity gradient $d{\log_{10}(Z_{eff}/Z_{\sun})}/dr$ 
in terms of the $K$ band disc scale length $h_K$
and the $K$ band central surface brightness, 
$K$ band absolute magnitude, $K$ band disc scale length and gas fraction.}
\label{fig:metgrad}
\end{minipage}
\end{figure*}

\begin{figure*}
\begin{minipage}{175mm}
\begin{center}
  \leavevmode
  \epsffile{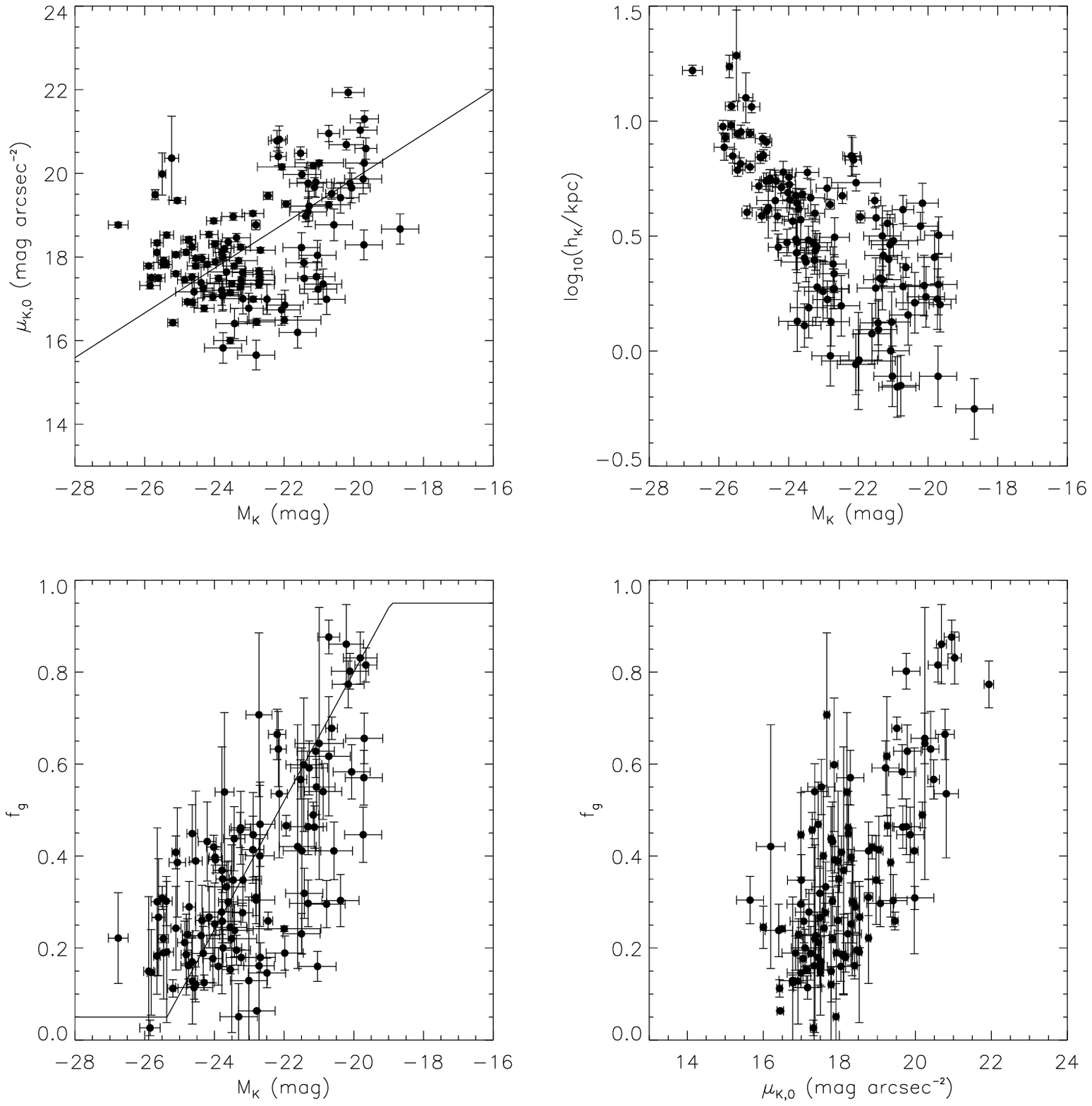}
\end{center}
\caption{  
Correlation between $K$ magnitudes,  
$K$ band central surface brightnesses, $K$ band disc scale lengths
and gas fractions.  Correlations
between the physical parameters are important to bear in mind when
interpreting the correlations in Figs.\ 
\protect{\ref{fig:age}}--\protect{\ref{fig:metgrad}}.
The solid line in the upper left panel 
is the unweighted least-squares bisector fit to the 
magnitude--central surface brightness correlation.
The solid line in the lower left panel is the estimated fit to the 
gas fraction--magnitude relation (see section \protect\ref{subsec:hii}).
}
\label{fig:phys}
\end{minipage}
\end{figure*}


In addition to the individual annulus age and metallicity estimates, we
calculated age and metallicity best-fit gradients and intercepts (as
described in section \ref{sec:grad}).  These fit parameters are
useful, in that they allow us to probe the {\it global} relationships between
e.g. magnitude and SFH.  

These fits to an individual galaxy are parameterised by their gradient
(in terms of the inverse 
$K$ band scale lengths) and their intercept
(expressed as the intercept at the disc half
light radius $R_{eff}$).  We explore correlations between the
structural parameters and the radial age and metallicity fit parameters
in Figs.\,\ref{fig:age}--\ref{fig:metgrad}.

\subsubsection{Global ages}

In Fig.\,\ref{fig:age}, we see how the age intercept at the 
half light radius relates to the $K$ band surface brightness, 
$K$ band absolute magnitude, $K$ band disc scale length and
gas fraction.  

As expected from the strong
correlation between the local surface brightness and age, there
is a highly significant correlation between the 
$K$ band central surface brightness and age.  We see also that there are 
also highly significant correlations between the $K$ band
absolute magnitude and age, and the gas fraction and age.  
There is no significant correlation between age and $K$ band disc
scale length.

An important question to ask at this stage is: are these correlations
related to each other in some way?  We investigate this in Fig.\
\ref{fig:phys}, where we plot some of the physical parameters against 
each other.  We see that the $K$ band absolute magnitude of this 
sample is correlated (with large scatter) with the 
$K$ band central surface brightness.  We also see in Fig.\ 
\ref{fig:phys} that the brightest galaxies all have the largest scale
lengths.  In addition, both magnitude and 
surface brightness correlate strongly with the gas fraction.
From these correlations, we can see that 
all of the three correlations 
between age and central surface brightness, magnitude and gas fraction
may have a common origin:  the correlation 
between these three physical parameters means that it
is difficult, using Fig.\,\ref{fig:age} alone, to determine
which parameters age is sensitive to.
We do feel, however, that the correlations
between age and surface brightness and age and magnitude
do not stem from the correlation between age and gas fraction for
two reasons.
Firstly, the scatter in the age--gas fraction
relationship is comparable to or even larger than the age--surface brightness
and age--magnitude correlations.  
Secondly, the gas fraction is an indication of the total 
amount of star formation over the history
of a galaxy (assuming that there is little late gas infall), implying
that the gas fraction is quite dependent on the SFH.  Therefore, 
correlations between SFH and gas fraction are likely to be driven
by this effect, rather than indicating that gas fraction drives SFH.

To summarise Figs.\,\ref{fig:age} and \ref{fig:phys}, 
the age of a galaxy is correlated strongly with the $K$ band 
surface brightness, $K$ band absolute magnitude and the gas fraction.
We discuss these important correlations further in section 
\ref{subsec:sbvsmg}.

\subsubsection{Global metallicities} \label{subsubsec:met}

In Fig.\,\ref{fig:met}, we see how the metallicity intercept at the 
disc half light radius relates to the $K$ band surface brightness, 
$K$ band absolute magnitude, $K$ band disc scale length and
gas fraction.  

The trends shown in Fig.\,\ref{fig:met} are similar to those seen 
in Fig.\,\ref{fig:age}:  this demonstrates the close correlation
between the age and metallicity of a galaxy, in the sense that 
older galaxies are typically much more metal rich than their younger 
brethren.  However, there are some important differences between the 
two sets of correlations.

Firstly, although the correlation is not statistically significant,
there is a conspicuous lack of galaxies with large scale lengths and 
low metallicities.  This lack of large, metal poor galaxies can be 
understood from the relationship between magnitude and
scale length in Fig.\,\ref{fig:phys}:  large galaxies, because of their
size, are bright by default (even if they have low surface brightnesses;
Paper {\sc i}) and have near-solar metallicities.

Secondly, there seems to be a kind of `saturation' in the 
stellar metallicities that was not apparent for the mean ages.
It is particularly apparent in the correlation between
the metallicity and $K$ band absolute magnitude:  the metallicity of
galaxies with an absolute $K$ band magnitude of $-22$ is very similar 
to the metallicity of galaxies that are nearly 
40 times brighter, with an absolute $K$ band magnitude
of $-26$.  The metallicity of galaxies fainter than an absolute $K$ 
band magnitude of $-22$ can be much lower, to the point where the metallicities
are too low to be included reliably on the stellar population model grid.
This `saturation' is easily understood in terms of the gas fractions:
the near-solar metallicity galaxies tend to have gas fractions $\la$ 0.5.
In this case, assuming a closed box 
model of chemical evolution (i.e.\ no gas flows into or out of
the galaxy, and the metals in the galaxy are well-mixed),
the stellar metallicity will be within 0.5 dex of the yield 
(see Pagel 1998 and references therein), 
which in this case would indicate
a yield around, or just above, solar metallicity.  
This case is shown in the lower right panel of Fig.\,\ref{fig:met}
(also Fig.\,\ref{fig:metcomp}), 
where we show the stellar metallicity of a solar metallicity
yield closed box model against the gas fraction.
Note that
the gas metallicity in the closed box model continues to rise to very
low gas fractions:  this will become important later when we
compare the stellar metallicity--luminosity relation from this work with 
the gas metallicity--luminosity relation (section \ref{subsec:hii}).

\subsubsection{Age gradients}

\begin{figure}
\begin{center}
  \leavevmode
  \epsffile{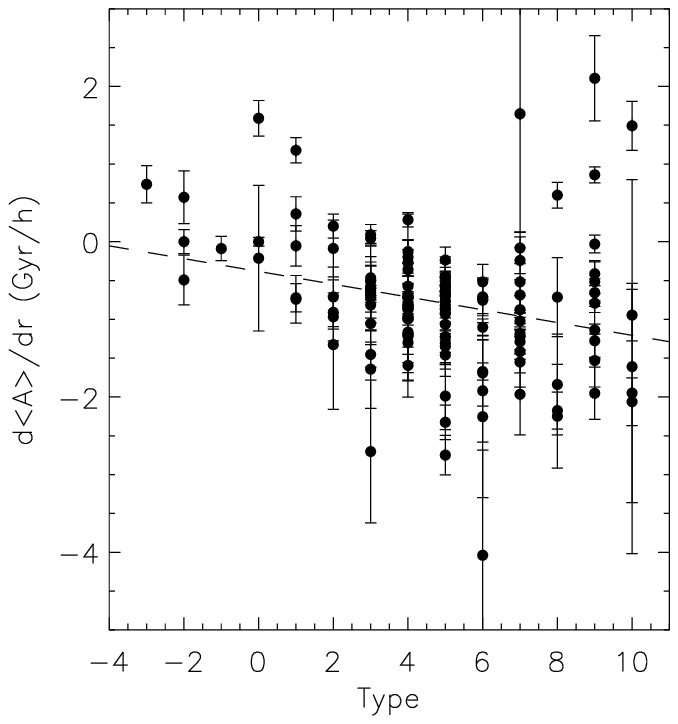}
\end{center}
\caption{Correlations between galaxy type and age gradient.
The x-axis is the T-type, where types between $-3$ and $-0$ 
correspond to S0s, and types 1 through 10 correspond to a
smooth sequence from Sa to Im.  The dashed line is the 
unweighted least-squares fit to the data. 
}
\label{fig:type}
\end{figure}

In Fig.\,\ref{fig:agegrad}, we see how the age gradient
(per $K$ band disc scale length) relates to the $K$ band surface brightness, 
$K$ band absolute magnitude, $K$ band disc scale length and
gas fraction.  

An important first result is that, on average, we have a 
strong detection of an overall age gradient:  the average
age gradient per $K$ band disc scale length is 
$-0.79 \pm 0.08$ Gyr\,$h^{-1}$ (10$\sigma$; where the quoted error is 
the error in the mean gradient).    

We see that the age gradient does not
correlate with the $K$ band absolute magnitude.
However, there are statistically significant correlations
between the age gradient and $K$ band central surface brightness, 
$K$ band disc scale length and gas fraction.
Smaller galaxies, higher surface brightness galaxies and
low gas fraction galaxies all tend to have relatively
flat age gradients (with much scatter, that almost exceeds the
amplitude of the trend).    

One possibility is that the S0s with `inverse' 
age gradients (in the sense that their central regions were
younger than their outer regions)  
are producing these trends, as the S0s in this sample are relatively
faint (therefore not producing much of a trend in age gradient with magnitude)
but have high surface brightness, small sizes and low gas fractions 
(therefore producing trends in all of the other physical parameters).

We investigate this possibility in Fig.\,\ref{fig:type}.  Here
we see that the trend is not simply due to `contamination' from 
a small population of S0s:  S0s contribute to a more general trend of 
decreasing age gradients for earlier galaxy types.  Therefore, this
decrease in age gradient is a real effect.   
We address a possible cause for this phenomenon 
later in section \ref{subsec:schmidt}.

\subsubsection{Metallicity gradients}

In Fig.\,\ref{fig:metgrad}, we see how the metallicity gradient
(per $K$ band disc scale length) relates to the $K$ band surface brightness, 
$K$ band absolute magnitude, $K$ band disc scale length and
gas fraction.  
The metallicity gradient is fairly poorly determined because of its 
typically smaller amplitude, and sensitivity to the rather noisier
near-IR colours.  Also, because of its smaller amplitude, 
metallicity gradients are more susceptible to the effects of dust reddening
than were the rather larger age gradients.  
On average (assuming that the effects of dust are negligible), 
we have detected   
an overall metallicity gradient (although with much scatter):  the average
metallicity gradient per $K$ band disc scale length is 
$-0.14 \pm 0.02$ dex\,$h^{-1}$ (7$\sigma$; where again the quoted error is 
the error in the mean gradient). 
Due to the large observational scatter, however, we have failed to detect
any trends in metallicity gradient with galaxy properties.


\begin{table*}
\begin{minipage}{135mm}
  \caption{Unweighted least-squares fits for the 
    statistically significant correlations (with
    significance
    $>$ 99 per cent)
    between SFH and galaxy structural parameters 
    presented in Figs. \protect\ref{fig:local}--\protect\ref{fig:metgrad}
    and \protect\ref{fig:type}--\protect\ref{fig:mod}.}
  \label{tab:fit}
  \begin{tabular}{ccr@{$\pm$}lr@{$\pm$}lr@{$\times$}l}
  \hline
   X & Y & \multicolumn{2}{c}{Slope} & \multicolumn{2}{c}{Intercept} & 
           \multicolumn{2}{c}{$P^a$} \\
  \hline
$\mu_{K}$ & $\langle A \rangle$ (Gyr) &     $-$0.42&     0.04 &
      8.20 &      0.07 &   6.2&10$^{-24}$ \\
$\mu_{K}$ & $\log_{10} (Z/Z_{\sun})$ &     $-$0.132 &     0.011
 &       $-$0.43 &      0.03 &   2.4&10$^{-30}$ \\
$\mu_{K,0}$ & $\langle A \rangle_{eff}$ (Gyr) &     $-$0.81 &      0.14 &
      6.6 &      0.3 &   2.9&10$^{-10}$ \\
$M_K$ & $\langle A \rangle_{eff}$ (Gyr) &     $-$0.38 &     0.09 &
    6.8 &       0.3 &   2.3&10$^{-6}$ \\
$f_g$ & $\langle A \rangle_{eff}$ (Gyr) &      $-$4.8 &      1.0 &
      9.6 &      0.3 &   4.1&10$^{-10}$ \\
$\mu_{K,0}$ & $\log_{10} (Z_{eff}/Z_{\sun})$ &     $-$0.22 &
    0.04 &       $-$0.80 &      0.09 &   4.9&10$^{-8}$ \\
$M_K$ & $\log_{10} (Z_{eff}/Z_{\sun})$ &     $-$0.15 &     0.02
 &      $-$0.86 &      0.09 &   2.6&10$^{-9}$ \\
$f_g$ & $\log_{10} (Z_{eff}/Z_{\sun})$ &      $-$1.4 &      0.3
 &      0.12 &     0.08 &   1.0&10$^{-8}$ \\
$\mu_{K,0}$ & $d\langle A \rangle/dr$ (Gyr/$h$) &     $-$0.25 &
    0.07 &       $-$1.2 &       0.2 &   1.4&10$^{-6}$ \\
$\log_{10}(h_K)$ & $d\langle A \rangle/dr$ (Gyr/$h$) &     $-$1.0 &   0.3
 &     $-$0.30 &      0.15 &   1.2&10$^{-6}$ \\
$f_g$ & $d\langle A \rangle/dr$ (Gyr/$h$) &      $-$1.1 &      0.5
 &     $-$0.5 &      0.2 &  4.5&10$^{-4}$  \\
Type & $d\langle A \rangle/dr$ (Gyr/$h$) &    $-$0.08 &     0.03 & 
    $-$0.38 &      0.16 &  5.9&10$^{-4}$  \\
$M_K$ & $\log_{10} (Z_{eff}/Z_{\sun}) - \log_{10} (Z_{\mu}/Z_{\sun})$ 
 &    -0.08 &     0.02 &     -0.22 &      0.07 & 1.9&10$^{-4}$ \\
$\mu_{K,0}$ & $\langle A \rangle_{eff} - \langle A \rangle_M$  (Gyr)
 &    -0.60 &      0.15 &       -0.9 & 0.3 &   1.3&10$^{-6}$ \\
$\mu_{K,0}$ & $\log_{10} (Z_{eff}/Z_{\sun}) - \log_{10} (Z_M/Z_{\sun})$ 
 &     -0.13 &     0.03 &  -0.22 &    0.08 &  2.6&10$^{-4}$  \\
Modified $\mu_{K,0}$ & $\langle A \rangle_{eff}$ (Gyr) &     -0.9 & 
     0.2 &       5.7 & 0.5 &   1.1&10$^{-7}$ \\
Modified $M_K$ & $\langle A \rangle_{eff}$ (Gyr) &     -0.52 & 0.09
 &       6.0 & 0.4 &   8.9&10$^{-9}$ \\
Modified $\mu_{K,0}$ & $\log_{10} (Z_{eff}/Z_{\sun})$ &     -0.21 &
 0.05 &     -0.9 &   0.1 &   4.9&10$^{-6}$ \\
Modified $M_K$ & $\log_{10} (Z_{eff}/Z_{\sun})$ &     -0.14 &
    0.03 &     -0.9 & 0.1 &   2.3&10$^{-7}$  \\
  \hline
  \end{tabular}
\\
$^a\,P$ is defined as the 
    probability of the correlation
    being the result of random fluctuations in an uncorrelated dataset:
    the usual definition of significance is $(1-P)$. \\
    Surface brightness intercepts are quoted at a $K$ band 
    surface brightness of 20 mag\,arcsec$^{-2}$. \\
    Absolute magnitude intercepts are quoted at a $K$ band 
    absolute magnitude of $-20$. \\
\medskip
\end{minipage}
\end{table*}

\section{Discussion} \label{sec:disc}

\subsection{Surface brightness vs.\ magnitude} \label{subsec:sbvsmg}
\begin{figure*}
\begin{minipage}{175mm}
\begin{center}
  \leavevmode
  \epsffile{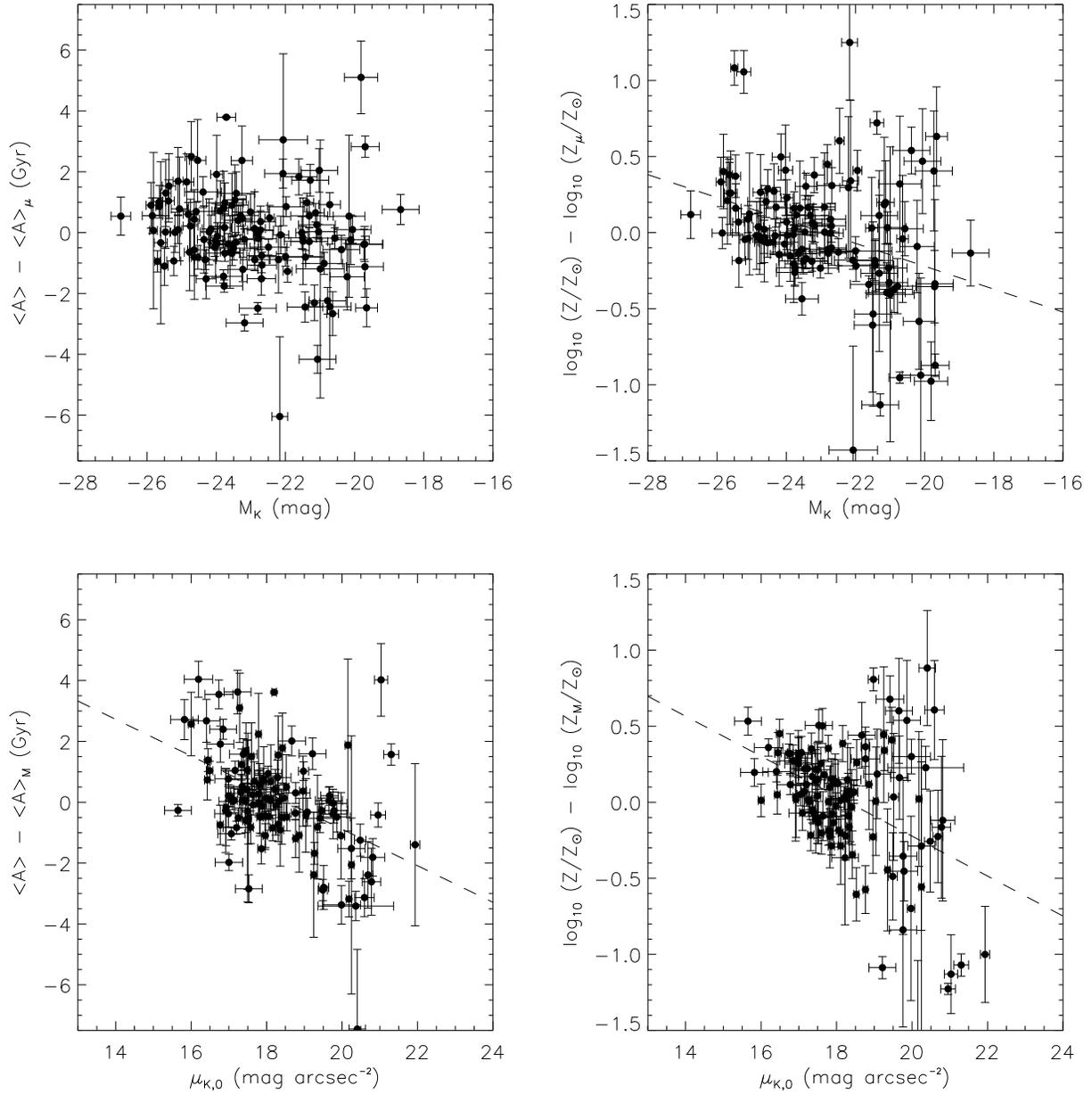}
\end{center}
\caption{
The top two panels show the residuals from the age--surface
brightness correlation of Fig.\,\protect\ref{fig:age}
and the metallicity--surface
brightness correlation of Fig.\,\protect\ref{fig:met}
against $K$ band absolute magnitude.
The lower two panels show the residuals from the age--magnitude
correlation of Fig.\,\protect\ref{fig:age}
and the metallicity--magnitude
correlation of Fig.\,\protect\ref{fig:met}
against $K$ band central surface brightness.
Dashed lines are unweighted least-squares fits to trends significant
at greater than the 99 per cent level (see Table \protect\ref{tab:fit}).
}
\label{fig:resid}
\end{minipage}
\end{figure*}

In the previous section, we presented a wide range of correlations between
diagnostics of the SFH and the physical parameters that
describe a spiral galaxy.
Our main finding was that
the SFH of a galaxy 
correlates well with both the $K$ band absolute magnitude
and $K$ band surface brightness.  However, this leaves 
us in an uncomfortable position:  because of the surface
brightness--magnitude correlation in our 
dataset (Fig.\,\ref{fig:phys}), it is impossible to tell from 
Figs.\,\ref{fig:age}, \ref{fig:met} and \ref{fig:phys} alone
what parameter is more important in determining the 
SFH of a galaxy.

One straightforward way to investigate which parameter
is more important is by checking if the offset of a galaxy 
from e.g.\ the age--magnitude correlation of Fig.\,\ref{fig:age}
correlates with surface brightness.  

In the upper panels
of Fig.\,\ref{fig:resid}, we show the residuals from the
age--surface brightness correlation and the 
metallicity--surface brightness correlation against
magnitude.  Age residual does not significantly 
correlate with magnitude (the correlation is
significant at only the 97 per cent level and is not shown).  Metallicity
residuals are significantly correlated with the magnitude, where brighter
galaxies tend to be more metal rich than expected from their surface brightness
alone.

In the lower panels of Fig.\,\ref{fig:resid}, we show the residuals from the
age--magnitude correlation and the metallicity--magnitude correlation.  
In contrast to the upper panels of this figure, there are highly significant
correlations between the age residual and surface brightness, and 
the metallicity residual and surface brightness.

Another way to consider the above is to investigate the distribution 
of galaxies in the three dimensional 
age--magnitude--surface brightness
and metallicity--magnitude--surface brightness
spaces.  Unweighted least-squares fits of the age and metallicity as a 
function of surface brightness and magnitude yield the surfaces:
\begin{eqnarray*}
\langle A \rangle_{eff} & = & 6.24(\pm0.14) - 0.71(\pm0.06) 
(\mu_{K,0} - 20)  \\  &  &  - 0.15(\pm0.03) (M_K + 20) \\
\log_{10} (Z_{eff}/Z_{\sun})  & =  & -0.99(\pm0.05)  \\ 
&  &  - 0.17(\pm0.01) (\mu_{K,0} - 20) \\ & & - 0.10(\pm0.01) (M_K + 20), 
\end{eqnarray*}
where the quoted errors (in brackets) are 
derived using bootstrap resampling of the data and the intercept
is defined at $\mu_{K,0} = 20$ mag\,arcsec$^{-2}$ and 
$M_K = -20$.
A best fit line to the dataset is shown in Figs.\,\ref{fig:age}
and \ref{fig:met}.   This best fit line is defined by the intersection of the
best-fit plane with the least-squares bisector fit 
(where each variable is treated equally in the fit; Isobe et al.\ 1990)
of the magnitude--surface brightness correlation (Fig.\,\ref{fig:phys}):
$M_K = -19.75 + 1.88(\mu_{K,0} - 20)$.
No plots of the distribution of galaxies in these three-dimensional
spaces are shown:  the two-dimensional projections of this space 
in Figs.\,\ref{fig:age}, \ref{fig:met} and \ref{fig:phys} are
among the best projections of this three-dimensional space.  

From the above fits, we can see quite clearly that 
age is primarily sensitive to central surface brightness:
the change in average age 
per magnitude change in central surface brightness 
is much larger than change in age per 
magnitude change in luminosity.  The stellar metallicity
of a galaxy is sensitive to both surface brightness and 
magnitude:  this is clear
from the fairly comparable change in metallicity per magnitude
change in surface brightness or luminosity.

We have also analysed the residuals from the age--local $K$ band
surface brightness and metallicity--local $K$ band surface brightness
correlations as a function of central $K$ band surface brightness
and $K$ band absolute magnitude (Fig.\,\ref{fig:local}).  
{\it The age of an annulus in a galaxy
primarily correlates with its local $K$ band surface brightness} and 
correlates to a lesser extent with the galaxy's $K$ band central 
surface brightness and magnitude.  {\it The metallicity
of a galaxy annulus correlates equally well with both the local
$K$ band surface brightness and total galaxy magnitude} and 
only very weakly with
the central $K$ band surface brightness.  Thus, the results
from the analysis of local ages and metallicities 
are consistent with the analysis of the global correlations:
much (but not all) 
of the variation that was attributed to the {\it central}
surface brightness in the global analysis is driven by 
the {\it local} surface brightness in the local analysis.
However, note that it is impossible to tell whether it 
is the global correlations that drive the local correlations, or vice
versa.  The local $K$ band surface brightnesses (obviously) correlate
with the central $K$ band surface brightnesses, meaning that it
is practically impossible to disentangle the effects of 
local and global surface brightnesses.

However, there must be other factors at play in determining the 
SFH of spiral galaxies:  the local and
global age and metallicity residuals 
(once all the above trends have been removed)
show a scatter around 1.4 times larger than the 
(typically quite generous) observational errors,
indicating cosmic scatter equivalent to $\sim$ 1 Gyr in age and $\sim$ 0.2
dex in metallicity.  This scatter is intrinsic to the galaxies 
themselves and {\it must} be explained in any model
that hopes to accurately describe the driving forces behind
the SFHs of spiral galaxies.  However, note that some or all of this
scatter may be due to the influence of small bursts of star formation 
on the colours of galaxies.

To summarise, {\it $K$ band surface brightness
(in either its local or global form)
is the most important `predictor' of the the SFH
of a galaxy}:  the effects of $K$ band absolute
magnitude modulate the overall trend defined by surface brightness.
Furthermore, it is apparent that the metallicity of a galaxy depends
more sensitively on its $K$ band magnitude than does the age: this point
is discussed later in section \ref{subsec:schmidt}.  On top of these 
overall trends, there is an intrinsic scatter, indicating that
the surface brightness and magnitude cannot be the only parameters
describing the SFH of a galaxy.

\subsection{Star formation histories in terms of masses and densities} 
  \label{subsec:mod}

\begin{figure*}
\begin{minipage}{175mm}
\begin{center}
  \leavevmode
  \epsffile{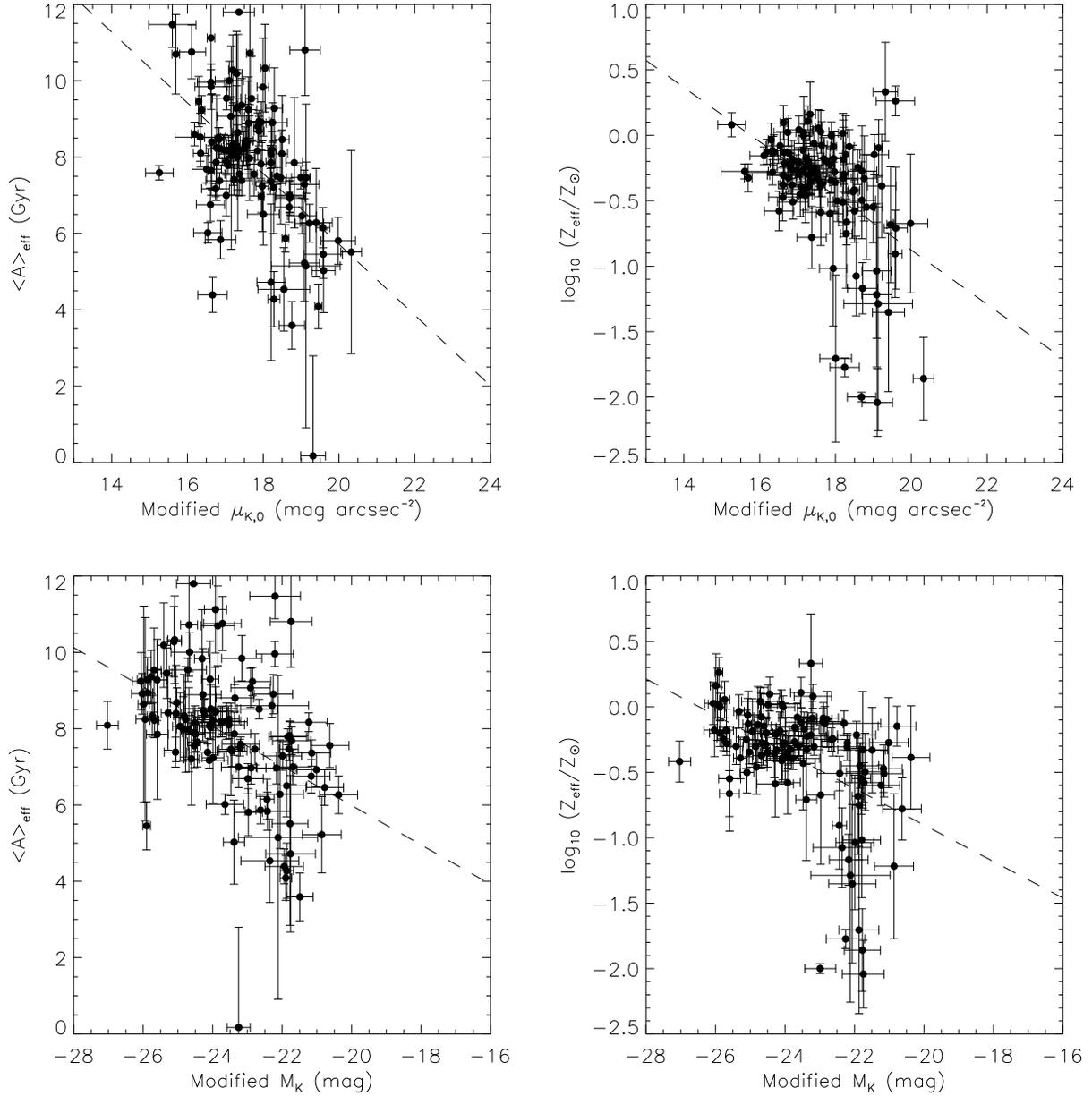}
\end{center}
\caption{
Correlations between the average age 
and metallicity 
at the disc half
light radius and the modified $K$ band central surface brightness 
and absolute magnitudes.  The surface brightnesses and absolute
magnitude are modified to approximate {\it total} baryonic densities and
masses, by turning the gas fraction of the galaxies into stars with 
a $K$ band mass to light ratio of 0.6 $M_{\sun}/L_{\sun}$.
Dashed lines are unweighted least-squares fits to trends significant
at greater than the 99 per cent level (see Table \protect\ref{tab:fit}).
}
\label{fig:mod}
\end{minipage}
\end{figure*}

In the above, we have seen how age and metallicity vary as a function
of $K$ band surface brightness and $K$ band absolute magnitude.  
$K$ band stellar mass to light ratios are expected to be 
relatively insensitive to differences in stellar populations (compared
to optical passbands; e.g.\ dJ{\sc iv}).  Therefore, the variation 
of SFH with $K$ band
surface brightness and absolute magnitude is likely to 
approximately represent
variations in SFH with stellar surface density
and stellar mass (especially over the dynamic range in surface brightness
and magnitude considered in this paper).

However, these trends with $K$ band surface brightness and 
$K$ band absolute magnitude may not accurately represent the 
variation of SFH with {\it total} (baryonic) surface density or mass:  
in order to take that into account, we must somehow account 
for the gas fraction of the galaxy.  Accordingly, we have 
constructed `modified' $K$ band surface brightnesses and
magnitudes by adding $2.5 \log_{10} (1 - f_g)$.  This correction, 
in essence, converts all of the gas fraction (both the measured atomic
gas fraction, and the much more uncertain molecular gas fraction) into
stars with a constant stellar mass to light ratio in $K$ band of 
0.6 $M_{\sun}/L_{\sun}$.  

In this correction, we make two main assumptions.  Firstly, 
we assume a constant $K$ band stellar mass 
to light ratio of 0.6 $M_{\sun}/L_{\sun}$.
This might not be such an inaccurate assumption:
the $K$ band mass to light ratio is expected to be relatively 
robust to the presence of young stellar populations; however,
our assumption of a constant $K$ band mass to light ratio is
still a crude assumption and should be treated with caution.
Note, however, that the relative trends in Fig.\,\ref{fig:mod}
are quite robust
to changes in stellar mass to light ratio: as the stellar mass to light ratio
increases, the modified magnitudes and surface brightnesses 
creep closer to their unmodified values asymptotically. 

Secondly, the correction
to the surface brightness 
implicitly assumes that
the gas will turn into stars with the same spatial distribution 
as the present-day stellar component.  This is quite a poor
assumption as the gas distribution is usually much more
extended than the stellar distribution:  this would imply that
the corrections to the central surface brightnesses are likely
to be overestimates.  However, our purpose here is not to construct
accurate measures of the baryonic masses and densities of disc galaxies; 
our purpose is merely to try to construct some kind of representative
mass and density which will allow us to determine if the trends 
in SFH with $K$ band magnitude and surface brightness reflect 
an underlying, more physically meaningful trend in SFH with
mass and density.

In Fig.\,\ref{fig:mod} we show the trends in 
age at the disc half light radius (left hand panels) and 
metallicity at the disc half light radius (right hand panels) with
modified $K$ band central surface brightness (upper panels) and 
modified $K$ band magnitude (lower panels).  {\it It is clear that
the trends in SFH with $K$ band surface brightness
and absolute magnitude presented in Figs.\,\ref{fig:age}
and \ref{fig:met} represent an underlying trend in 
SFH with the total baryonic galaxy densities and masses.}
The best fit slopes are typically slightly
steeper than those for the unmodified magnitudes and surface 
brightnesses (Table \ref{tab:fit}):  
because low surface brightness and/or faint
galaxies are gas-rich, correcting them for the contributions
from their gas fractions tends to steepen the correlations somewhat. 
We have also attempted to disentangle surface density and mass
dependencies in the SFH using the method described above in 
section \ref{subsec:sbvsmg}:  again, we find that surface 
density is the dominant parameter in 
determining the SFH of a galaxy, and that
the mass of a galaxy has a secondary, modulating effect on the SFH.

\subsection{Does the choice of IMF and SPS model affect our conclusions?}
  \label{subsec:imf}

We have derived the above conclusions using a Salpeter \shortcite{sp}
IMF and the BC98 models, but are our conclusions affected by 
SPS model details such as IMF, or the choice of
model?   We address this possible source of
systematic uncertainty in Table \ref{tab:imf}, where we compare the 
results for the correlation between the $K$ band central surface
brightness $\mu_{K,0}$ and age intercept at the disc half
light radius $\langle A \rangle_{eff}$ using the models of BC98 with
the IMF adopted by Kennicutt \shortcite{ke} and the models of 
KA97 with a Salpeter IMF.  

\begin{table}
  \caption{Unweighted least-squares fits for the 
    correlation between the $K$ band central surface
    brightness $\mu_{K,0}$ and age intercept at the half light 
    radius $\langle A \rangle_{eff}$ using different SPS models and IMFs. }
  \label{tab:imf}
  \begin{tabular}{ccr@{$\pm$}lr@{$\pm$}lr@{$\times$}l}
  \hline
   Model & IMF & \multicolumn{2}{c}{Slope} & \multicolumn{2}{c}{Intercept} & 
           \multicolumn{2}{c}{$P^a$} \\
  \hline
   BC98 & Salpeter  & $-$0.81 & 0.14 & 6.6 & 0.3 &  2.9 & 10$^{-10}$ \\    
   BC98 & Kennicutt & $-$0.93 & 0.14 & 6.2 & 0.3 &  3.3 & 10$^{-10}$ \\ 
   KA97 & Salpeter  & $-$0.82 & 0.12 & 6.1 & 0.3 &  8.8 & 10$^{-11}$ \\   
  \hline   
  \end{tabular} \\ $^a\,P$ is defined as the 
    probability of the correlation
    being the result of random fluctuations in an uncorrelated dataset:
    the usual definition of significance is $(1-P)$. \\
    The intercept is quoted at a $K$ band central 
    surface brightness of 20 mag\,arcsec$^{-2}$.
  \medskip
\end{table}

From Table \ref{tab:imf}, it is apparent that our results are quite
robust to changes in both the stellar IMF and SPS model used to interpret
the data.  
In none of the cases does the significance of the
correlation vary by a significant amount, and the
slope and intercept of the correlation varies within its estimated
bootstrap resampling errors.  This correlation is not exceptional; 
other correlations show similar behaviour, with little variation
for plausible changes in IMF and SPS model.
In conclusion, while plausible changes in the IMF and SPS model 
will change the details of the correlations (e.g.\ the 
gradients, intercepts, significances, etc.), the existence
of the correlations themselves is very robust to changes in
the models used to derive the SFHs.  

\subsection{How important is dust extinction?}
  \label{subsec:dust}

Our results account for average age and metallicity 
effects only; however, from Figs.\,\ref{fig:colcoltype}--\ref{fig:colcolfg}
it is clear that dust reddening can cause colour changes similar to those
caused by age or metallicity.
In this section, we discuss
the colour changes caused by dust reddening.  We conclude that
dust reddening will mainly affect the metallicity 
(and to a lesser extent age) gradients:  however, the magnitude
of dust effects is likely to be too small to significantly affect our
conclusions. 

In Figs.\,\ref{fig:colcoltype}--\ref{fig:colcolfg}, we 
show the reddening vectors for two different
dust models (Milky Way and Small Magellanic Cloud dust models)
and for two different reddening geometries (a simple foreground 
screen model and exponential star and dust disc model).  
We can see that the reddening effects of the foreground
screen (for a total $V$ band extinction of 0.3 mag) are
qualitatively similar (despite its unphysical dust geometry) 
to the central $V$ band absorption 
of a $\tau_V = 2$ Triplex dust model.  
For the Triplex model, the length of the
vector shows the colour difference between the central 
(filled circle) and
outer colours, and the open circle denotes the colour effects at the disc
half light radius.

How realistic are these Triplex vectors?  dJ{\sc iv} compares
absorption-only face-on Triplex dust reddening vectors with 
the results of Monte Carlo simulations, including the effects of scattering,
concluding that the Triplex model vectors are quite accurate for
high optical depths but that they considerably {\it overestimate} the 
reddening for lower optical depths (i.e.\ $\tau_V \la 2$).
However, both these models include only the effects of smoothly 
distributed dust.  If a fraction of the dust is distributed in 
small, dense clumps, the reddening produced by a given overall dust mass
decreases even more:  the dense clumps of dust tend to simply `drill
holes' in the face-on galaxy light distribution, producing very little
reddening per unit dust mass \cite{hu94,djiv,k98}.
The bottom line is that in such a situation, the galaxy colours
are dominated by the least obscured stars, with the dense
dust clouds having little effect on the overall colour.
Therefore, the Triplex model vectors in Figs.\ 
\ref{fig:colcoltype}--\ref{fig:colcolfg} are arguably overestimates
of the likely effects of dust reddening on our data.

From Figs.\,\ref{fig:colcoltype}--\ref{fig:colcolfg}, we
can see that the main effect of dust reddening would be the 
production of a colour gradient which would mimic small 
artificial age and metallicity gradients.  
Note, however, that the amplitudes of the majority of the observed 
colour gradients
are larger than the Triplex model vectors.
In addition, we have checked for trends in age and metallicity
gradient with galaxy ellipticity: no significant trend was found, suggesting
that age and metallicity trends are unlikely to be solely 
produced by dust.
Coupled with the above arguments,
this strongly suggests that most of the colour gradient of a given galaxy 
is likely to be due to stellar population differences.
This is consistent with the findings of Kuchinski et al.\ \shortcite{k98}, 
who found, using more realistic dust models tuned to reproduce accurately
the colour trends in high-inclination galaxies, that the colour
gradients in face-on galaxies were likely to be due primarily to 
changes in the underlying stellar populations with radius.
Therefore, our measurements of the age and metallicity gradients 
are likely to be qualitatively correct, but trends in 
dust extinction with e.g.\ magnitude or surface brightness 
may cause (or hide) weak trends in the gradients.

The age and metallicity intercepts
at the half light radius would be relatively unaffected:
in particular, differences between the central $V$ band optical depths 
of 10 or more 
cannot produce trends in the ages or metallicities with
anywhere near the dynamic range observed in the data.
{\it Therefore, the main conclusion of this paper, that the 
SFH of a spiral galaxy is primarily driven by surface density and
is modulated by mass, is robust to the effects of dust reddening}.

\subsection{Comparison with \hii region metallicities}
  \label{subsec:hii}


\begin{figure*}
\begin{minipage}{175mm}
\begin{center}
  \leavevmode
  \epsffile{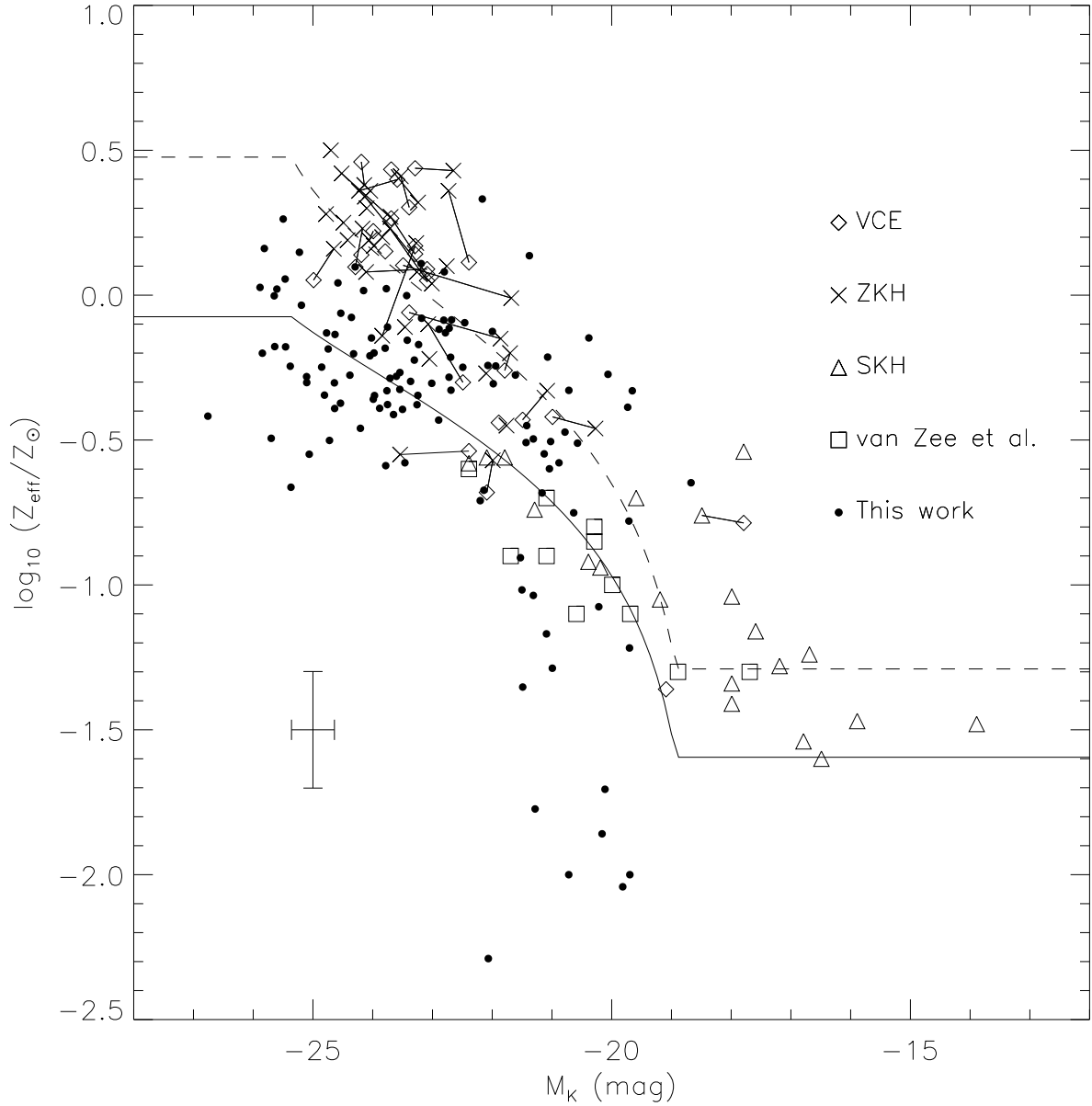}
\end{center}
\caption{  
Comparison between the trends in gas metallicity and magnitude 
and stellar metallicity and magnitude.  Our data from Fig. 
\protect \ref{fig:met} are overplotted on \protect \hii region 
metallicity determinations at a radius of 3 kpc for the sample of 
Vila-Costas \& Edmunds (1992; VCE)
(diamonds) and Zaritsky et al.\ (1994; ZKH)
(diagonal crosses).  Global gas metallicity determinations for dwarf 
galaxies are taken from Skillman et al.\ (1989; SKH)
(triangles) and van Zee et al.\ \protect \shortcite{vanzee1997} (squares).
Multiple metallicity determinations for the same galaxy are 
connected by solid lines.  The dashed and solid lines are the 
closed box predictions for the gas metallicity--magnitude  and 
stellar metallicity--magnitude relations, respectively (see text for details).
}
\label{fig:metcomp}
\end{minipage}
\end{figure*}

In Fig. \ref{fig:metcomp}, we
plot our colour-based stellar metallicities at the disc 
half light radius against
measures of the global gas metallicity via \hii region spectroscopy.
$B$ band magnitudes were transformed into $K$ band magnitudes using an 
average $B-K$ colour of $3.4 \pm 0.4$ \cite{djiv}.
For bright galaxies, we have plotted metallicity determinations 
from Vila-Costas \& Edmunds
\shortcite{ve92} (diamonds) and Zaritsky et al.\ 
\shortcite{zkh94} (diagonal crosses) at a fixed radius
of 3 kpc.  To increase our dynamic range in terms of galaxy magnitude,
we have added global gas metallicity measures from the studies of 
Skillman et al.\ \shortcite{skh89} and van Zee et al.\ 
\shortcite{vanzee1997}.  Measurements for the same galaxies in different
studies are connected by solid lines.

From Fig. \ref{fig:metcomp}, it is clear that our colour-based stellar
metallicities are in broad agreement with the trends in gas metallicity
with magnitude explored by the above \hii region studies.  
However, there are two notable differences between the colour-based
stellar metallicities and the \hii region metallicities.  

Firstly, there is a `saturation' in stellar metallicity at bright
magnitudes not seen in the gas metallicities, which continue to
rise to the brightest magnitudes.
In section \ref{subsubsec:met}, 
we argued that this `saturation' was due to the 
slow variation of stellar metallicity with gas fraction 
at gas fractions lower than $\sim$ 1/2.
In Fig.\,\ref{fig:metcomp} we test this idea.
The dashed line is the gas metallicity--magnitude relation 
expected if galaxies evolve as closed boxes with solar
metallicity yield, converting between gas fraction and 
magnitude using the 
by-eye fit to the magnitude--gas fraction correlation
$f_g = 0.8 + 0.14(M_K - 20)$ (where the gas fraction
$f_g$ is not allowed to drop below 0.05 or rise above 0.95;  
see Fig.\,\ref{fig:phys}).  The solid 
line is the corresponding relation for the stellar metallicity.
Note that at $K$ band absolute magnitudes brighter than 
$-$25 or fainter than $-$19 the model metallicity--magnitude
relation is poorly constrained:  the gas fractions at the bright and
faint end asymptotically approach zero and unity respectively. 
The closed box model indicates that our interpretation of 
the offset between gas and stellar metallicities at the brightest
magnitudes is essentially correct:  the gas metallicity of galaxies 
is around 0.4 dex higher than the average stellar metallicity.  Furthermore, 
the slope of the gas metallicity--magnitude relation is slightly
steeper than the stellar metallicity--magnitude relation.
However, the simple closed box model is far from perfect:
this model underpredicts the 
stellar metallicity of spiral galaxies with $K \sim -21$.  This
may be a genuine shortcoming of the closed box model; however,
note that we have crudely translated gas fraction into magnitude: 
the closed box model 
agrees with observations quite closely if 
we plot gas and stellar metallicity against gas fraction 
(Fig.\,\ref{fig:met}).

Secondly, there is a sharp drop in the estimated stellar metallicity
at faint magnitudes that is not apparent in the gas metallicities.  
This drop, which occurs for stellar metallicities lower than $\sim 1/10$
solar, is unlikely to be physical:  stellar and gas metallicities
should be quite similar at high gas fractions (equivalent to faint 
magnitudes; Fig.\,\ref{fig:metcomp}).  
However, the SPS
models are quite uncertain at such low 
metallicities, suggesting that the SPS models
overpredict the near-IR magnitudes of very metal-poor
composite stellar populations.

We have also compared our derived metallicity gradients with the gas 
metallicity gradients from Vila-Costas \& Edmunds \shortcite{ve92}
and Zaritsky et al.\ \shortcite{zkh94}.  Though our measurements have
large scatter, we detect an average metallicity 
gradient of $-0.06 \pm 0.01$ kpc$^{-1}$ for the whole sample.
This gradient is quite comparable to the average metallicity gradient
from the studies
Vila-Costas \& Edmunds \shortcite{ve92}
and Zaritsky et al.\ \shortcite{zkh94} of
$-0.065 \pm 0.007$ kpc$^{-1}$.
Given the simplicity of the 
assumptions going into the colour-based analysis, and considerable SPS
model and dust reddening uncertainties, we take the broad agreement
between the gas and stellar metallicities as an important confirmation 
of the overall validity of the colour-based method. 


\subsection{A possible physical interpretation}
  \label{subsec:schmidt}


\begin{figure*}
\begin{minipage}{175mm}
\begin{center}
  \leavevmode
  \epsffile{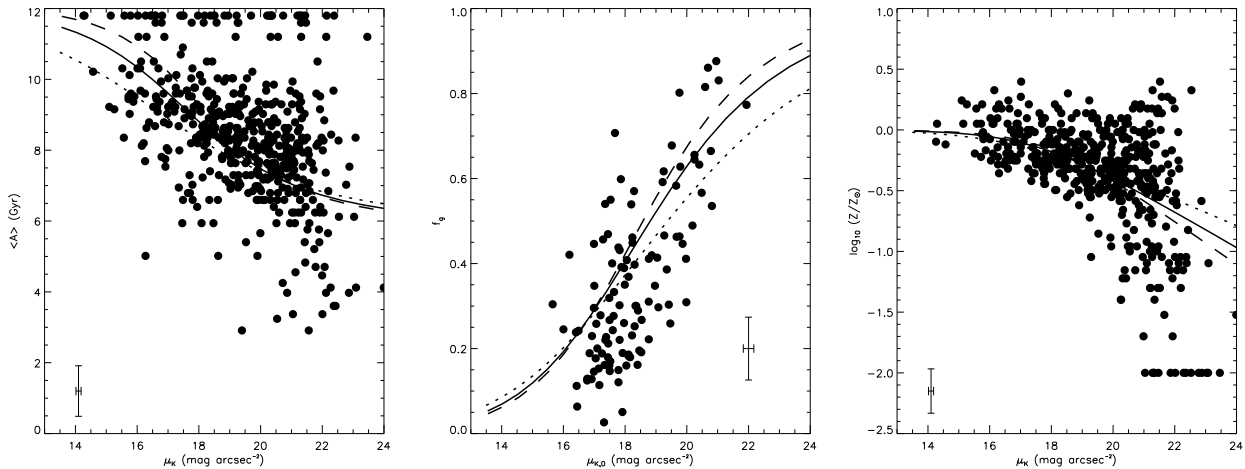}
\end{center}
\caption{  
Local average age (left), gas fraction (middle) 
and metallicity (right) against $K$ band
surface brightness.  The lines are predictions of 
the local density dependent model presented in the text with
$n = 1.4, k = 0.020$ (dotted), $n = 1.6, k = 0.007$ (solid) and
$n = 1.8, k = 0.0025$ (dashed).
}
\label{fig:schmidt}
\end{minipage}
\end{figure*}

In order to demonstrate the utility of the trends
presented in section \ref{sec:res} in investigating star formation 
laws and galaxy evolution, we consider a simple model of galaxy evolution.
We have found that the surface density of a galaxy is the most
important parameter in describing its SFH; therefore,
we consider a simple model where the star formation 
and chemical enrichment history are controlled by
the surface density of gas in a given region
\cite{schmidt1959,phillipps1990,phillipps1991,dopita1994,elm94,prantzos1998,ken98}.
In our model, we assume that the initial state 
12 Gyr ago is a thin layer of gas
with a given surface density $\sigma_g$.  We allow this layer of
gas to form stars according to a Schmidt \shortcite{schmidt1959} star
formation law:
\begin{equation} \label{eqn:schmidt}
\Psi(t) = -\frac{1}{\alpha} \frac{d\sigma_g(t)}{dt} = k \sigma_g(t)^n,
\end{equation}
where $\Psi(t)$ is the SFR at a time $t$, $\alpha$
is the fraction of gas that is locked up in long-lived stars 
(we assume $\alpha = 0.7$ hereafter; this value is appropriate
for a Salpeter IMF), 
$k$ is the efficiency of star formation and the exponent $n$ determines
how sensitively the star formation rate depends on the gas surface
density \cite{phillipps1991}.
Integration of Equation \ref{eqn:schmidt} gives:
\begin{equation} \label{eqn:gasfracneq1}
\sigma_g(t)/\sigma_0 = e^{-\alpha k t},
\end{equation}
for $n=1$ and:
\begin{equation} \label{eqn:gasfracnne1}
\sigma_g(t)/\sigma_0 = [(n-1) \sigma_0^{n-1} \alpha k t + 1 ]^{1/1-n},
\end{equation}
for $n \ne 1$ where $\sigma_0$ is the initial surface 
density of gas.  
Note that the 
star formation (and gas depletion) history of the $n=1$ case is
{\it independent} of initial surface density.  To be consistent with 
the trends in SFH with surface brightness presented in this paper 
$n$ must be larger than one.

We allow no gas inflow or outflow from this region, i.e. our model 
is a closed box.  In addition, in keeping with our definition of the 
star formation law above (where we allow a fraction of the gas consumed 
in star formation to be returned immediately, enriched with 
heavy elements), we use the instantaneous recycling approximation.
While this clearly will introduce some inaccuracy into our model
predictions, given the simplicity of this model a more sophisticated
approach is not warranted.  Using these approximations, it is possible
to relate the stellar metallicity $Z$ to the gas fraction $f_g$
(e.g.\ Pagel \& Pratchett 1975; Phillips \& Edmunds 1991; Pagel 1998):
\begin{equation}
Z(t) = p (1+\frac{f_g \log f_g}{1-f_g}),
\end{equation}
where $p$ is the yield, in this case fixed at solar metallicity $p = 0.02$.
Numerical integration of the SFR from  
Equations \ref{eqn:schmidt}, \ref{eqn:gasfracneq1} and 
\ref{eqn:gasfracnne1} allows determination of the average age of the 
stellar population $\langle A \rangle$.  

In modelling the SFH in this way, we are attempting to describe
the broad strokes of the trends in SFH observed in Fig. \ref{fig:local}, 
using the variation in SFH caused by variations in {\it initial
surface density alone}.  In order to meaningfully compare the models to the
data, however, it is necessary to translate the initial surface density
into a surface brightness.  This is achieved using the 
solar metallicity SPS models of 
BC98 to translate the model SFR into a $K$ band surface brightness. 
While the use of the solar metallicity model ignores the effects 
of metallicity in determining the $K$ band surface brightness,
the uncertainties in the $K$ band model mass to light ratio are sufficiently
large that tracking the metallicity evolution of the $K$ band surface
brightness is unwarranted for our purposes.  

Using the above models, we tried to reproduce the trends in Fig. 
\ref{fig:local} by adjusting the values of $n$ and $k$.  We found that 
reasonably good fits to both trends are easily achieved by balancing off 
changes in $n$ against compensating changes in $k$ over quite
a broad range in $n$ and $k$.  In order to improve
this situation, we used the 
correlation between $K$ band central surface brightness 
and gas fraction as an additional constraint:  in order to predict
the relationship between central surface brightness and global gas fraction, 
we constructed the mass-weighted gas fraction for an 
exponential disc galaxy with the
appropriate $K$ band central surface brightness.  
We found that models with $n \sim 1.6$ and
$k \sim 0.007$ fit the observed trends well, with modest increases in $n$
being made possible by decreasing $k$, and vice versa.  Plots of the fits
to the age, metallicity and gas fraction trends with surface brightness
are given in Fig. \ref{fig:schmidt}.

The success of this simple, local 
density-dependent model in reproducing the broad trends 
observed in Fig. \ref{fig:schmidt} is quite compelling, re-affirming
our assertion that {\it surface density plays a dominant r\^{o}le in 
driving the SFH and chemical enrichment
history of spiral galaxies}.
This simple model also explains the origin of age 
and metallicity gradients
in spiral galaxies:  the local surface density in spiral galaxies
decreases with radius, leading to younger ages and lower
metallicities at larger radii.  Indeed, the simple model
even explains the trend in age gradient with surface brightness:
high surface brightness galaxies will have a smaller age gradient per
disc scale length because of the flatter slope of the (curved)
age--surface brightness relation at high surface brightnesses.

However, our simple model is clearly inadequate:
there is no explicit mass dependence in this
model, which is required by the data.
This may be alleviated by the use of 
a different star formation law.  Kennicutt \shortcite{ken98}
studied the correlation between global gas surface density and SFR,
finding that a Schmidt law with exponent 1.5 {\it or} a
star formation law which depends on both local gas surface density
and the dynamical timescale (which depends primarily on the rotation
curve shape, and, therefore, on mostly global parameters) explained
the data equally well.
There may also be a surface density threshold below which 
star formation cannot occur \cite{ken89}.
In addition, the fact that galaxy metallicity depends on magnitude
and surface brightness in almost equal amounts (age is much more
sensitive to surface brightness) suggests that e.g.\ galaxy
mass-dependent feedback may be an important process in the 
chemical evolution of galaxies.  
Moreover, a closed box chemical evolution model
is strongly disfavoured by e.g.\ studies of the metallicity distribution 
of stars in the Milky Way, where closed box models hugely overpredict
the number of low-metallicity stars in the solar neighbourhood and 
other, nearby galaxies \cite{worthey1996,pagel}.  
This discrepancy can be solved by allowing gas to flow in and out
of a region, and indeed this is expected in 
galaxy formation and evolution models set in the 
context of large scale structure formation 
\cite{cole1994,kauffmann1998,val99}.  

\section{Conclusions} \label{sec:conc}

We have used a diverse sample of low-inclination spiral galaxies with 
radially-resolved optical and near-IR photometry to investigate
trends in SFH with radius, as a function of 
galaxy structural parameters.  A maximum-likelihood analysis
was performed, comparing SPS model colours
with those of our sample galaxies, allowing use of all of the 
available colour information.  Uncertainties in the 
assumed grid of SFHs, the SPS models and uncertainties due to dust reddening
were not taken into account.  Because of these uncertainties,
the absolute ages and metallicities we derived may be inaccurate;
however, our conclusions will be robust in a relative sense.
In particular, dust will mainly affect the age and metallicity gradients; 
however, the majority of a given galaxy's age or metallicity gradient
is likely to be due to gradients in its stellar population alone. 
The global age and metallicity trends are robust to the effects of 
dust reddening.  Our main conclusions are as follows.
\begin{itemize}
\item Most spiral galaxies have stellar population  gradients, in the sense
that their inner regions are older and more metal
rich than their outer regions.  The amplitude of 
age gradients increase from high surface brightness to low
surface brightness galaxies.  An exception to this trend are
faint S0s from the Ursa Major Cluster of galaxies:  the central
stellar populations of these galaxies are younger and more metal
rich than the outer regions of these galaxies.
\item The stellar metallicity--magnitude relation `saturates' for 
the brightest galaxies.  This `saturation' is a real effect:  as the
gas is depleted in the brightest galaxies, the gas metallicity tends
to rise continually, while the stellar metallicity 
flattens off as the metallicity tends towards the yield.  
The colour-based metallicities of the faintest spirals fall considerably 
($\ga 1$ dex) below the gas metallicity--luminosity relation:  this 
may indicate that the SPS models overpredict the 
$K$ band luminosity of very low metallicity composite stellar populations.
\item There is a strong correlation between the SFH
of a galaxy (as probed by its age, metallicity and gas fraction)
and the $K$ band surface brightness and magnitude of that galaxy.  From 
consideration of the distribution of galaxies in 
the age--magnitude--surface brightness and 
metallicity--magnitude--surface brightness spaces, we find that 
the SFH of a galaxy correlates primarily with either its local or global
$K$ band surface brightness:  the effects of $K$ band absolute
magnitude are of secondary importance. 
\item When the gas fraction is taken into account, the correlation
between SFH and surface density remains, with a small
amount of mass dependence.  Motivated by the strong correlation 
between SFH and surface density, and by the 
correlation between age and local $K$ band surface brightness, 
we tested the observations against a closed box local density-dependent
star formation law.  We found that despite its simplicity, 
many of the correlations
could be reproduced by this model, {\it indicating that the 
local surface density is the most important parameter in shaping
the SFH of a given region in a galaxy}.  
A particularly significant
shortcoming of this model is the lack of a magnitude dependence 
for the stellar metallicity:  this magnitude dependence 
may indicate that {\it mass-dependent feedback is an 
important process in shaping the chemical evolution of a galaxy}.
However, there is significant cosmic scatter in these correlations
(some of which may be due to small bursts of recent star formation),
suggesting that the mass and density of a galaxy may not be the 
only parameters affecting its SFH.  
\end{itemize}

\section*{Acknowledgements}

We wish to thank Richard Bower for his comments on early versions of
the manuscript and Mike Edmunds, Harald Kuntschner and Bernard 
Rauscher for useful discussions.  
EFB would like to thank the Isle of
Man Education Board for their generous funding.
Support for RSdJ was provided by NASA through Hubble Fellowship
grant \#HF-01106.01-98A from the Space Telescope Science Institute,
which is operated by the Association of Universities for Research in
Astronomy, Inc., under NASA contract NAS5-26555.  
This project made use of STARLINK facilities in Durham.
This research has made use of the NASA/IPAC Extragalactic Database (NED)
which is operated by the Jet Propulsion Laboratory, California Institute of
Technology, under contract with the National Aeronautics and Space
Administration.

\end{document}